%% file: CERN-EP-2024-309.tex
\pdfoutput=1
\documentclass[aps,prl,nofootinbib,twocolumn,showpacs,superscriptaddress,letterpaper,amsmath,amsfonts,amssymb,preprintnumbers,10pt]{revtex4-2}

\usepackage[bookmarks=false]{hyperref}
\usepackage{graphicx}
\usepackage{bm}
\usepackage{xspace}
\usepackage[separate-uncertainty=true]{siunitx}
\usepackage[italic]{hepnames}
\usepackage{xcolor}
\usepackage{orcidlink}
\usepackage{cleveref}
\usepackage{textcomp}
\usepackage{rotating}
\usepackage{booktabs}
\usepackage{array}
\usepackage{float}
\usepackage{multirow}
\usepackage{placeins}

\graphicspath{{figures/}}

\DeclareSIUnit{\barn}{b}
\DeclareSIUnit{\fb}{\femto\barn}
\DeclareSIUnit{\ifb}{\per\fb}
\DeclareSIUnit{\mm}{\milli\meter}
\DeclareSIUnit{\cm}{\centi\meter}

\crefname{appendix}{Appendix}{Appendices}
\crefname{equation}{Eq.}{Eqs.}
\crefname{figure}{Fig.}{Figs.}
\crefname{table}{Table}{Tables}

\providecommand{\papersection[1]}{\paragraph{#1}}

\creflabelformat{equation}{#2\textup{#1}#3}

\begin{document}

\title{
  First Measurement of the Muon Neutrino Interaction Cross Section and Flux as a Function of Energy at the LHC with FASER\\
  \vspace*{0.08in}
  {\normalsize FASER Collaboration}
}

\input{authorlist}

\begin{abstract}
  This letter presents the measurement of the energy-dependent neutrino-nucleon cross section in tungsten and the differential flux of muon neutrinos and anti-neutrinos. The analysis is performed using proton-proton collision data at a center-of-mass energy of \SI{13.6}{\TeV} and corresponding to an integrated luminosity of \qty{65.6 \pm 1.4}{\ifb}. Using the active electronic components of the FASER detector, \num{338.1 +- 21.0} charged current muon neutrino interaction events are identified, with backgrounds from other processes subtracted. We unfold the neutrino events into a fiducial volume corresponding to the sensitive regions of the FASER detector and interpret the results in two ways: We use the expected neutrino flux to measure the cross section, and we use the predicted cross section to measure the neutrino flux. Both results are presented in six bins of neutrino energy, achieving the first differential measurement in the TeV range. The observed distributions align with Standard Model predictions. Using this differential data, we extract the contributions of neutrinos from pion and kaon decays.
\end{abstract}

\preprint{CERN-EP-2024-309}

\date{\today}

\maketitle

\onecolumngrid

\begin{center}
  \textcopyright \ 2024 CERN for the benefit of the FASER Collaboration. Reproduction of this article or parts of it is allowed as specified in the CC-BY-4.0 license.
\end{center}

\twocolumngrid

\papersection{Introduction}
The Large Hadron Collider (LHC) produces an intense beam of neutrinos in the forward direction, originating from the decay of pions, kaons, and charmed hadrons from proton-proton ($pp$) collisions. The precision study of these neutrinos will carry broad implications for investigating neutrino properties, quantum chromodynamics, astroparticle physics, and exploring phenomena beyond the Standard Model~\cite{abreuDetectingStudyingHighenergy2020,Feng:2022inv}. In particular, they provide an opportunity to measure the neutrino interaction cross section in a previously unexplored energy range spanning from \qty{360}{\GeV} to \qty{6.3}{\TeV}, bridging the gap between fixed-target measurements~\cite{navasReviewParticlePhysics2024} and astroparticle data~\cite{Aartsen:2017kpd}. Studying collider neutrinos was originally suggested in 1984~\cite{DeRujula:1984pg}, but these neutrinos have only been measured recently, since they are outside of the instrumented acceptance of typical LHC experiments. In spring 2023, the FASER Collaboration directly observed collider neutrinos for the first time~\cite{FASER:2023zcr}. This observation was shortly after confirmed by the SND@LHC Collaboration~\cite{SNDLHC:2023pun}. Neutrino interaction cross sections at \unit{\TeV} energies were measured for the first time using the FASER emulsion detector~\cite{FASER:2024hoe}. These measurements mark collectively the dawn of collider neutrino physics~\cite{Worcester:2023njy}. In this letter, we report the first measurement of the charged current (CC) interaction cross section and the differential flux of muon neutrinos and anti-neutrinos as a function of the neutrino energy. The analysis uses $pp$ collision data recorded by the FASER detector and corresponding to an integrated luminosity of \qty{65.6 +- 1.4}{\ifb} collected at a center-of-mass energy of \qty{13.6}{\TeV}.

\papersection{The FASER Detector}
The FASER detector is located in the TI12 tunnel, approximately \qty{480}{\m} downstream of the ATLAS interaction point (IP1) and is aligned with the beam collision axis line-of-sight (LOS). Due to the $pp$ crossing angle in IP1, the LOS is shifted downward by about \qty{6.5}{\cm} with respect to the center of the detector\footnote{The $pp$ half-crossing angle in IP1 was \qty{-160}{\micro\radian} for 2022 data taking and varied between \qty{-165}{\micro\radian} and \qty{-135}{\micro\radian} for 2023 data taking resulting in slightly different LOS positions.}. The detector is shielded from the ATLAS IP by about \qty{100}{\m} of rock and concrete. Most background particles are thus either absorbed or are deflected by the LHC magnets, whereas weakly interacting neutrinos are unaffected.

The detector is briefly described below, with a more detailed overview provided in Ref.~\cite{abreuFASERDetector2024}. FASER consists of a FASER$\nu$ emulsion neutrino detector, the VetoNu, Veto, Timing and Pre-shower scintillator stations, a tracking spectrometer including three \qty{0.57}{\tesla} dipole magnets, and an electromagnetic calorimeter. The active transverse area of the detector is defined by the \qty{200}{\mm} diameter magnet aperture. The FASER$\nu$ emulsion detector consists of \num{730} layers of interleaved tungsten plates and emulsion films. It has a width of \qty{25}{\cm}, a height of \qty{30}{\cm} and a total mass of \num{1.1} metric tons. In this analysis, FASER$\nu$ serves only as the target for neutrino interactions and the electronic detector components are used to identify the neutrino events. The VetoNu scintillator station is positioned in front of the neutrino target in order to veto any incoming charged particles. It consists of two modules measuring $\qtyproduct{30 x 35}{\cm}$, which is significantly larger than the target and therefore allows charged particles entering with an angle with respect to the LOS to also be discarded. The other scintillator stations are positioned after the neutrino target and are used for triggering and timing measurements. The Veto scintillator station is located after the FASER$\nu$ emulsion detector and before the first magnet, the Timing station is positioned between the first and second magnets, and the Pre-shower station is located after the last magnet and in front of the calorimeter. Events can be triggered by signals from the scintillator stations and calorimeter, with a typical trigger rate of about \qty{1}{\kilo\Hz}~\cite{FASER:2021cpr}. The tracking system consists of an interface tracking station (IFT) and three tracking spectrometer stations~\cite{FASER:2021ljd}. Each tracking station is constructed from three planes comprising eight double-sided ATLAS semiconductor tracker (SCT) barrel modules~\cite{Abdesselam:2006wt} per plane.

\papersection{Dataset and Simulated Samples}
This analysis uses data obtained during runs with stable beam conditions collected in 2022 and 2023. These periods correspond to a total integrated luminosity of \qty{65.6 +- 1.4}{\ifb}~\cite{ATLAS:2022hro,atlas:2023lumi,atlas:2024lumi} after data quality selection. To study the detector response to neutrino interactions, \num{2.8e6} charged and neutral current (NC) neutrino interactions corresponding to an integrated luminosity of \qty{10}{\per\atto\barn} were simulated. The neutrino fluxes were obtained using the fast neutrino flux simulation of Ref.~\cite{Kling:2021gos} and adjusted to match the LHC Run~3 configuration~\cite{fasercollaborationNeutrinoRatePredictions2024}. \texttt{EPOS-LHC}~\cite{Pierog:2013ria} is used to simulate the production of light hadrons, and the predictions from \texttt{POWHEG}~\cite{Nason:2004rx,Frixione:2007vw,Alioli:2010xd} and \texttt{PYTHIA~8.3}~\cite{Bierlich:2022pfr} are employed to model charmed hadron production~\cite{buonocorePredictionsNeutrinosNew2024}. The interaction of neutrinos with the detector is simulated using the \texttt{GENIE~3.04.0}~\cite{Andreopoulos:2009rq,Andreopoulos:2015wxa,GENIE:2021zuu} event generator. All other particle interactions in the FASER detector are simulated using \texttt{GEANT4}~\cite{GEANT4:2002zbu}, with the \texttt{FTFP\_BERT} physics list as the default. FASER's offline software system is based on the Athena software from the ATLAS Collaboration~\cite{ATL-PHYS-PUB-2009-011} and the ACTS software framework~\cite{Ai:2021ghi} is used for the reconstruction of charged particle tracks.

\papersection{Selection and Background Rejection}
The analysis is optimized to select CC muon neutrino interactions within the fiducial volume, which produce a high-momentum muon traversing the entire detector. The fiducial volume is defined as a cylinder with a diameter of \qty{200}{\mm} around the central axis of the tracking spectrometer, closely resembling the experimental setup. The analysis selections and background estimates were finalized prior to examining data in the signal region to prevent bias. Events are selected if triggered by any scintillator downstream of the VetoNu scintillators, which operate with full efficiency for muons. A colliding bunch crossing at the ATLAS interaction point is required to exclude beam backgrounds and cosmic muons. We use the measured charge in the VetoNu scintillator station to discard charged background events. Neutrino interactions in the target can produce a charge in the VetoNu scintillator predominantly from low-energy neutrons produced through the de-excitation of tungsten nuclei. Thus, we additionally use the time difference between the VetoNu scintillators in front of the target and the Veto scintillators after the target to identify neutrino interactions. For background events, this time difference corresponds to the muon's time of flight. In contrast, neutrino interactions produce charge in the two scintillator stations from different particles, typically leading to larger time differences. This is exploited by defining a reduced charge as the measured charge in each VetoNu scintillator, integrated over the time window expected for a background muon to deposit energy. The reduced charge closely matches the total scintillator charge for muons but is typically much smaller for neutrinos. We require it to be less than \qty{30}{\pico\coulomb}, ensuring high signal efficiency while rejecting almost all background events. We also require scintillator signals compatible with a muon in all downstream scintillator layers. We require at least one track that passes through all three tracking stations of the FASER spectrometer. For events with multiple tracks the one with the highest momentum is assumed to originate from the muon and is used to determine the neutrino energy. A large number of secondary particles can be produced in neutrino interactions, which can saturate the IFT tracking station located directly after the FASER$\nu$ emulsion detector. As a result, only the three tracking spectrometer stations are used for track reconstruction. To exclude events with tracks that do not traverse the full spectrometer or have poor quality, tracks are required to have at least 14 hits in the silicon tracker across at least seven layers\footnote{A good track passing through all tracking stations typically has 18 silicon hits across nine layers} and to possess a reasonable fit quality. The tracks are reconstructed starting from the most downstream tracking station in order to increase the efficiency for selecting CC neutrino interactions, which can have a large number of hits in the upstream tracking stations. To reduce the contribution from background events, the track is required to have momentum greater than \qty{100}{\GeV}. It is extrapolated to estimate its position at the Veto and VetoNu  scintillator stations. We require that the track must lie within \qty{95}{\mm} from the central axis of the detector in all tracking stations, including the IFT, and within \qty{120}{\mm} at the VetoNu scintillator station. Finally, we require that the angle of the track with respect to the detector axis is less than \qty{25}{\milli\radian}. These radial and angular cuts suppress incoming charged particles that enter FASER at large radii or angles. With these selections we expect to observe almost exclusively CC muon neutrino and anti-neutrino interactions.

\papersection{Background Estimation}
The primary source of backgrounds arises from high-momentum muons, originating upstream of FASER. Almost all of these are vetoed by the VetoNu scintillators. The inefficiencies of the two layers were measured to be smaller than $10^{-7}$ per layer using events which pass the event selection but have a reduced charge larger than \qty{30}{\pico\coulomb} in one of the two layers. Assuming that the two scintillator layers are independent, we expect for the recorded luminosity about $10^{-6}$ muons going undetected through both layers, which is negligible compared to the other backgrounds. Muons incident at a large radius and low momentum can geometrically evade the VetoNu scintillator station but may still create a signal-like track in the detector. The expected yield is estimated from data using a sideband containing tracks that pass the event selection but have a momentum smaller than \qty{100}{\GeV}. As the contribution from neutrinos is non-negligible in the higher momentum bins of the sideband, we subtract the expected neutrino yields from the data. We then fit the track momentum distribution in the range of \qtyrange{10}{100}{\GeV} and extrapolate to our signal region. The extrapolation is performed separately for positively and negatively charged tracks to estimate the background in charge and momentum bins. The resulting number of geometric background (Geo. bgr.) events is compatible with zero in all bins. More details on the geometric background are given in appendix~\ref{app:geo_bkg}. Muons may also interact with material upstream of the detector and produce secondary neutral hadrons that can enter the FASER detector and produce tracks. We estimate the contribution of the neutral hadron background by using simulations corresponding to about \qty{2}{\per\atto\barn} of integrated luminosity within a radial distance of \qty{150}{\mm}. A large fraction of these events fail the event selection, either because the parent muon, which generated the neutral hadron, hits the VetoNu scintillator and fails the reduced VetoNu charge cut, or because the neutral hadron interacts and is fully absorbed in the 8 interaction lengths of the FASER$\nu$ emulsion detector, without leaving any track through the detector. With this, the expected number of neutral hadron events, passing the event selection, is less than $10^{-3}$ events for the studied integrated luminosity, and is therefore considered negligible. The backgrounds from cosmic rays and LHC beam background have been studied using events occurring when there are no collisions, and are found to be negligible. In addition to the signal neutrino interactions, identified as CC muon neutrino interactions in the fiducial volume (\Pnum CC fid.), there are a small number of electron neutrino (\Pnue CC) or tau neutrino interactions (\Pnut CC), interactions outside of the fiducial volume (\Pnum CC non-fid.), and neutral current neutrino interactions (\Pnu NC) that pass the event selection. We estimate the contribution of these neutrino interactions from simulation and subtract them from data. The resulting number of events is listed in \cref{tab:num-signal-events-qop}.

\begin{table*}[tb]
  \centering
  \caption{Number of simulated signal and background, and observed events  for an integrated luminosity of \qty{65.6}{\ifb}\ for the six $q/p^{\prime}_{\Pmu}$ bins, with $q$ denoting the muon unit charge and $p^{\prime}_{\Pmu}$ the calibrated muon momentum. The uncertainty on the simulated number of neutrino interactions is dominated by the flux uncertainty.}
  \label{tab:num-signal-events-qop}
  \begin{ruledtabular}
    \begin{tabular}{clS[table-format=2.1(1.1)]S[table-format=2.1(1.1)]S[table-format=2.1(1.1)]S[table-format=2.1(1.1)]S[table-format=2.1(1.1)]S[table-format=2.1(1.1)]S[table-format=3.1(1.1)]}
      & $q/p^{\prime}_{\Pmu} [{\rm GeV^{-1}}]$ & {{$\left[\frac{-1}{100}, \frac{-1}{300}\right]$}} & {{$\left[\frac{-1}{300}, \frac{-1}{600}\right]$}} & {{$\left[\frac{-1}{600}, \frac{-1}{1000}\right]$}} & {{$\left[\frac{-1}{1000}, \frac{1}{1000}\right]$}} & {{$\left[\frac{1}{1000}, \frac{1}{300}\right]$}} & {{$\left[\frac{1}{300}, \frac{1}{100}\right]$}} & {Total} \\
      \midrule
      \multicolumn{9}{c}{\textbf{Simulation}}                                                                                     \\
      \midrule
      Signal & $\nu_{\mu}$ CC (fid.)     & 33.6+-7.6 & 59.5+-9.1  & 51.6+-8.8 & 84.1+-21.4 & 50.1+-11.4 & 19.6+-5.9 & 298.4+-42.6 \\
      \parbox[t]{4mm}{\multirow{5}{*}{\rotatebox[origin=r]{0}{Bgr. $\begin{cases}  \\ \\ \\ \\ \end{cases}$}}}
             & $\nu_{\mu}$ CC (non-fid.) & 3.6+-1.4  & 3.7+-1.7   & 2.3+-1.3  & 2.0+-1.2   & 2.6+-1.3   & 2.9+-1.3  & 17.1+-6.6   \\
             & $\nu_{e}$ CC              & 0.7+-0.5  & 0.2+-0.2   & 0.1+-0.1  & 0.1+-0.2   & 0.4+-0.5   & 0.9+-0.7  & 2.3+-1.9    \\
             & $\nu_{\tau}$ CC           & 0.0+-0.1  & 0.0+-0.1   & 0.0+-0.1  & 0.0+-0.0   & 0.1+-0.1   & 0.0+-0.1  & 0.2+-0.4    \\
             & $\nu$ NC                  & 1.4+-0.5  & 0.4+-0.4   & 0.1+-0.2  & 0.1+-0.2   & 0.6+-0.4   & 1.3+-0.8  & 4.0+-1.4    \\
             & Geo. bgr.                 & 0.2+-0.6  & 0.0+-0.1   & 0.0+-0.0  & 0.0+-0.0   & 0.0+-0.0   & 0.0+-0.0  & 0.3+-0.6    \\
      \midrule
             & Total                     & 39.5+-9.2 & 63.9+-10.2 & 54.1+-9.6 & 86.3+-22.3 & 53.7+-12.1 & 24.7+-7.5 & 322.3+-50.5 \\
      \midrule
      \multicolumn{9}{c}{\textbf{Data}}                                                                                           \\
      \midrule
             & Total                     & 50        & 97         & 71        & 69         & 48         & 27        & 362         \\
    \end{tabular}
  \end{ruledtabular}
\end{table*}

\papersection{Signal Extraction}
We apply a linear scaling to the reconstructed muon momentum, such that $p^{\prime}_{\Pmu} = p_{\Pmu} / 0.8$. The factor of 0.8 describes that muons passing the event selection typically retain, on average, \SI{80}{\percent} of the incident neutrino energy and results in a more diagonal response matrix, and therefore smaller uncertainties on the unfolded number of neutrino interactions. We define six bins in the ratio of the muon unit charge $q$ and calibrated muon momentum, $q/p^{\prime}_{\Pmu}$, with bin edges
\begin{align}
  \left[ -\frac{1}{100}, -\frac{1}{300}, -\frac{1}{600}, -\frac{1}{1000}, \frac{1}{1000}, \frac{1}{300}, \frac{1}{100} \right] \, \si{\per\GeV} \, . \nonumber
\end{align}
This binning results in a comparable expected number of events in each bin, and separates muons originating from neutrinos and anti-neutrinos up to a momentum of \qty{1}{\TeV}. The charge identification accuracy is significantly reduced above this threshold, and we thus combine muons and anti-muons with a larger momentum into a single bin.

We unfold the number of reconstructed muons ($n_{\Pmu}^i$) to the number of neutrino interactions in the fiducial volume ($n_{\Pnu}^k$) within bins of the ratio of the negative lepton number\footnote{Note that the negative lepton number $-L$ of the neutrino is equal to the unit charge $q$ of the muon produced in the CC neutrino interaction.} and neutrino energy, $-L/E_{\Pnu}$, via
\begin{equation}\label{eq:unfolding}
  n_{\Pmu}^i = \sum_k M_{ik} \, \epsilon_k \, n_{\Pnu}^k + n_{\rm bkg}^i \ ,
\end{equation}
with $M_{ik}$ denoting the response matrix, which relates the reconstructed, calibrated muon momentum in $q/p^{\prime}_{\Pmu}$ \mbox{bin $i$} to the neutrino energy in $-L/E_{\Pnu}$ bin $k$ (c.f. appendix~\ref{app:response_matrix}). Further, $\epsilon_k$ denotes the efficiency of neutrino events to be reconstructed within the fiducial volume, and $n_{\rm bkg}^i$ are the estimated backgrounds from geometric muons and non-signal neutrino interactions, which includes \Pnum CC interactions outside of the fiducial volume.

The response matrix and efficiency are obtained from simulation and systematic uncertainties are implemented with nuisance parameters. The number of neutrino interactions is determined through a binned extended maximum likelihood fit, expressed as:
\begin{align}\label{eq:likelihood}
  \mathcal{L} = \prod_i^{\rm bins} \mathcal{P}( N_{\Pmu}^i | n_{\Pmu}^i ) \times \prod_l \mathcal{G}_l \, ,
\end{align}
where $\mathcal{P}$ represents a Poisson distribution with $N_{\Pmu}^i$ as the number of observed events and $n_{\Pmu}^i$ as the number of expected events (calculated using \cref{eq:unfolding}) in bin $i$ of $q/p^{\prime}_{\Pmu}$. The term $\mathcal{G}_l$ corresponds to Gaussian priors constraining nuisance parameters from source $l$. The likelihood function in \cref{eq:likelihood} is numerically minimized using the \texttt{iminuit} package~\cite{iminuit}.

Using the simulated neutrino flux, we calculate the energy-dependent neutrino-nucleon interaction cross section $\sigma^k $ from the unfolded and efficiency corrected yields for each $-L/E_{\Pnu}$ bin $k$, via
\begin{align}
  \sigma^k = \frac{n_{\nu}^k}{   \rho_A \cdot \mathrm{L} \cdot \iint  j^k_{\mathrm{sim}} \, \mathrm{d} E \mathrm{d} A } \, ,
  \label{eq:cross-section}
\end{align}
where $\iint j^k_{\mathrm{sim}} \, \mathrm{d} E \mathrm{d} A$ denotes the simulated neutrino flux density $j_{\mathrm{sim}}$ integrated over the cross sectional area of the fiducial volume $A$ and the neutrino energy of a given bin $E$ in units of number of neutrinos per integrated luminosity, $\rho_A = \qty{1.022 +- 0.010e27}{\text{nucleon}{\per\cm\squared}}$ is the area nucleon density and $\mathrm{L} = \qty{65.6 +- 1.4}{\ifb}$ is the integrated luminosity.

Conversely, assuming the theoretical cross section, the total flux $\phi^k$ through the fiducial volume in each neutrino energy bin $k$ is determined via
\begin{equation}
  \phi^k = \frac{ n_{\nu}^k }{ \rho_A \cdot \mathrm{L} \cdot \sigma_{\rm sim}^k} \, .
\end{equation}
Since the neutrino cross section varies within a measured bin, we use the flux-weighted cross section $\sigma_{\rm sim}$ within a bin. This depends on how the flux varies in each energy bin, and we apply a systematic uncertainty related to the variation using the different generators considered.
Additionally, we follow the approach in Ref.~\cite{fasercollaborationNeutrinoRatePredictions2024} and apply a \qty{6}{\percent} uncertainty to account for the difference between cross section predictions of the Bodek-Yang model~\cite{Bodek:2002vp,Bodek:2004pc,Bodek:2010km} and more recent models~\cite{Candido:2023utz,Jeong:2023hwe}.

\begin{figure}[bth]
  \centering
  \includegraphics[width=\linewidth]{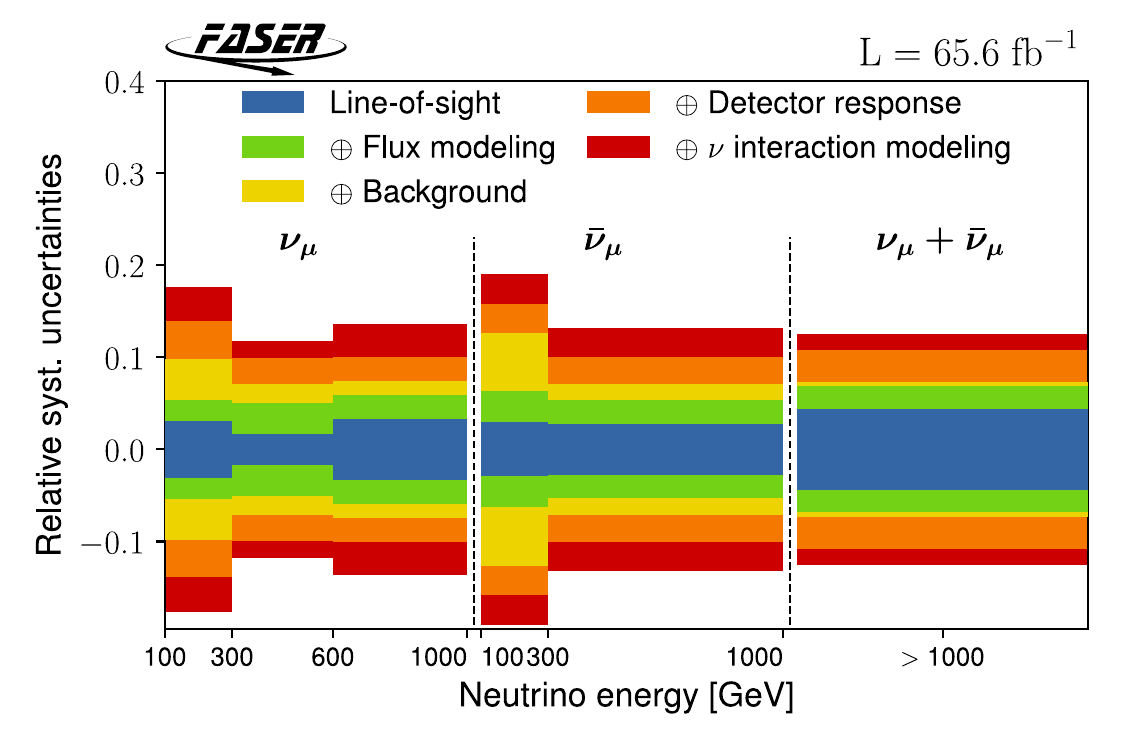}
  \caption{The relative systematic uncertainties on the determined number of fiducial neutrino interactions. The uncertainty components are added in quadrature such that the outer contour represents the total systematic uncertainty.}
  \label{fig:syst-uncertainties}
\end{figure}

\papersection{Systematic Uncertainties}
The presented results are affected by several systematic uncertainties. To estimate the uncertainty from the modeling of the neutrino interactions and the passage of the particles, produced in the neutrino interaction, through the detector, we run \texttt{GENIE} and \texttt{GEANT4} with different tunes and physics lists modifying the final state interactions and hadron transport. More details on these models are given in appendix~\ref{app:modeling_unc}. To quantify a systematic uncertainty due to mismodeling of detector components or responses in the simulation, we vary the nominal selection requirements within a range determined by comparisons between data and simulation. In particular, the positions of the SCT modules in the tracking stations can deviate from their nominal positions, resulting in a different momentum scale and resolution between simulation and data. Following a similar approach to that used in Ref.~\cite{CMS:2021ime}, we study the impact of these misalignments on our analysis using simulations with misaligned geometries that reproduce the performance observed in data. The systematic uncertainty on the number of background geometric muon events originates from the statistical uncertainties on the data and the uncertainty on the simulated number of neutrinos in the $p_{\mu}<\qty{100}{\GeV}$ region. The uncertainty on the number of non-signal neutrino interactions is estimated from simulation and dominated by the flux uncertainty. To estimate the uncertainty from neutrino flux predictions, we follow Ref.~\cite{fasercollaborationNeutrinoRatePredictions2024} and simulate neutrino interactions using the flux from the MC generators described in appendix~\ref{app:modeling_unc}. The neutrino rate and energy distribution depend on the distance from the LOS, which is determined by the beam crossing angle at the IP. We simulate neutrino interactions for half-crossing angles of both \qty{-160}{\micro\radian} and \qty{-135}{\micro\radian} and apply the difference as a systematic uncertainty for all time periods where the crossing angle was not \qty{-160}{\micro\radian} (\qty{26.9}{\percent} of all events). \Cref{fig:syst-uncertainties} summarizes the resulting relative systematic uncertainties on the determined number of fiducial neutrino interactions in bins of the neutrino energy.

\begin{figure}[bth]
  \centering
  \includegraphics[width=\linewidth]{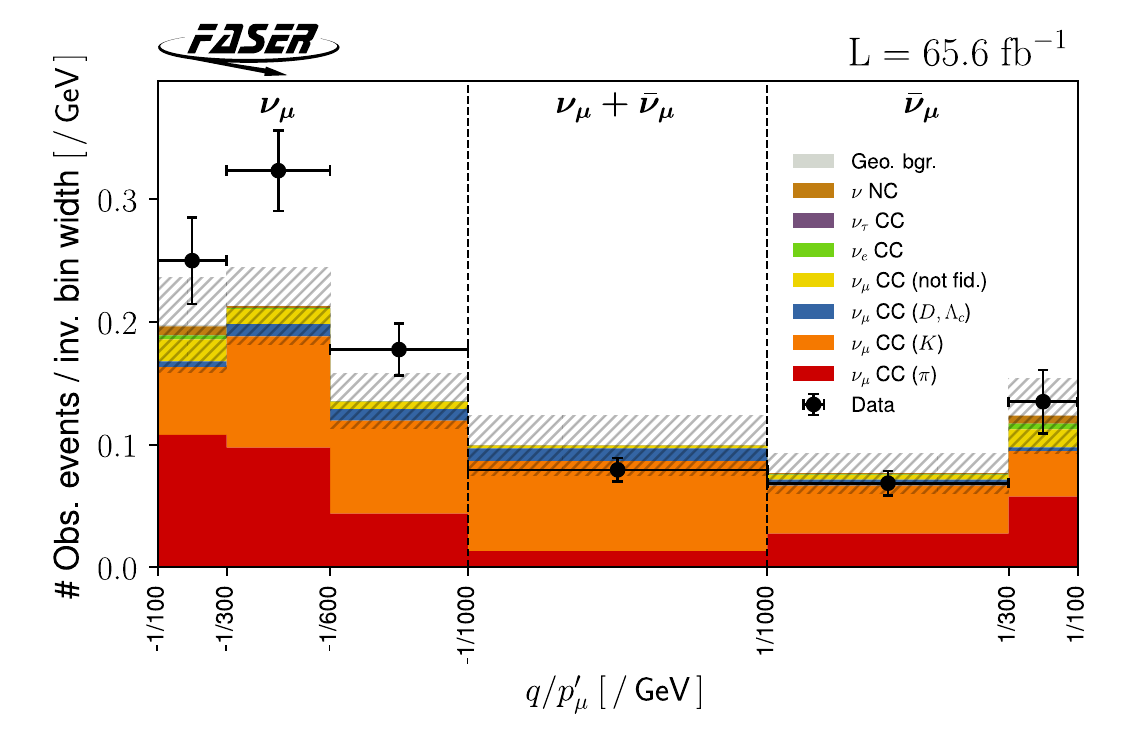}
  \includegraphics[width=\linewidth]{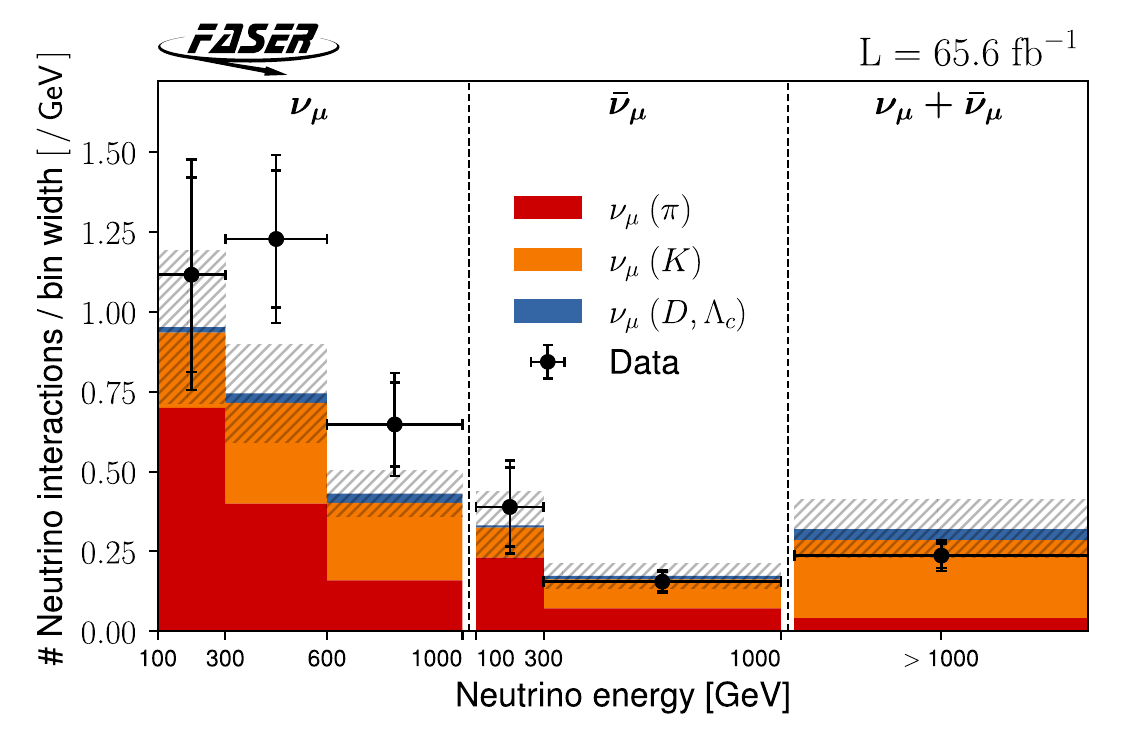}
  \vspace*{-0.15in}
  \caption{
    (\textbf{Top}) The number of observed neutrino candidate events per inverse bin width (black markers) are compared to the expected number of CC, NC, and background contributions (histogram).
    (\textbf{Bottom}) The unfolded number of neutrino interactions per bin width (black markers) is compared to the expectation from the simulation (histograms). For the common high-energy bin we use a width of \qty{868.2}{\GeV} which contains \SI{68.3}{\percent} of the simulated events. The inner error bars show the statistical uncertainty and the outer error bars the total uncertainty. The hatched area in both plots indicates the systematic uncertainty, whose dominant source stems from the neutrino flux predictions.
  }
  \label{fig:observed_events}
\end{figure}

\papersection{Results}
After subtracting the number of expected background events, we observe
\begin{align}
  n_{\Pnu,{\rm obs}} = \num[parse-numbers=false]{338.1 \pm 19.0 \, ({\rm stat.}) \pm 8.8 \, ({\rm sys.})}
\end{align}
events from CC muon neutrino and anti-neutrino interactions. This agrees well with the expected value of \num{298.4+-42.6} within the uncertainties. \Cref{tab:num-signal-events-qop} and \cref{fig:observed_events} (top panel) show the number of observed and expected signal events in each bin. We observe more negatively charged muons with an energy between \qty{300}{\GeV} and \qty{1}{\TeV} than expected from simulation. More properties of the selected neutrino candidate events are summarized in appendix~\ref{app:sig_bkg_comp}.

Using the likelihood fit, defined in ~\cref{eq:likelihood}, and the efficiency and response matrix, detailed in appendix~\ref{app:response_matrix}, we determine $n_{\Pnu,{\rm fid}} = 1242.7 \pm 137.1$ CC muon neutrino interactions inside the fiducial volume. \cref{fig:observed_events} (bottom panel) shows the unfolded number of neutrino interactions and compares them to the expectation from simulation.

\begin{figure}[bth]
  \centering
  \includegraphics[width=\linewidth]{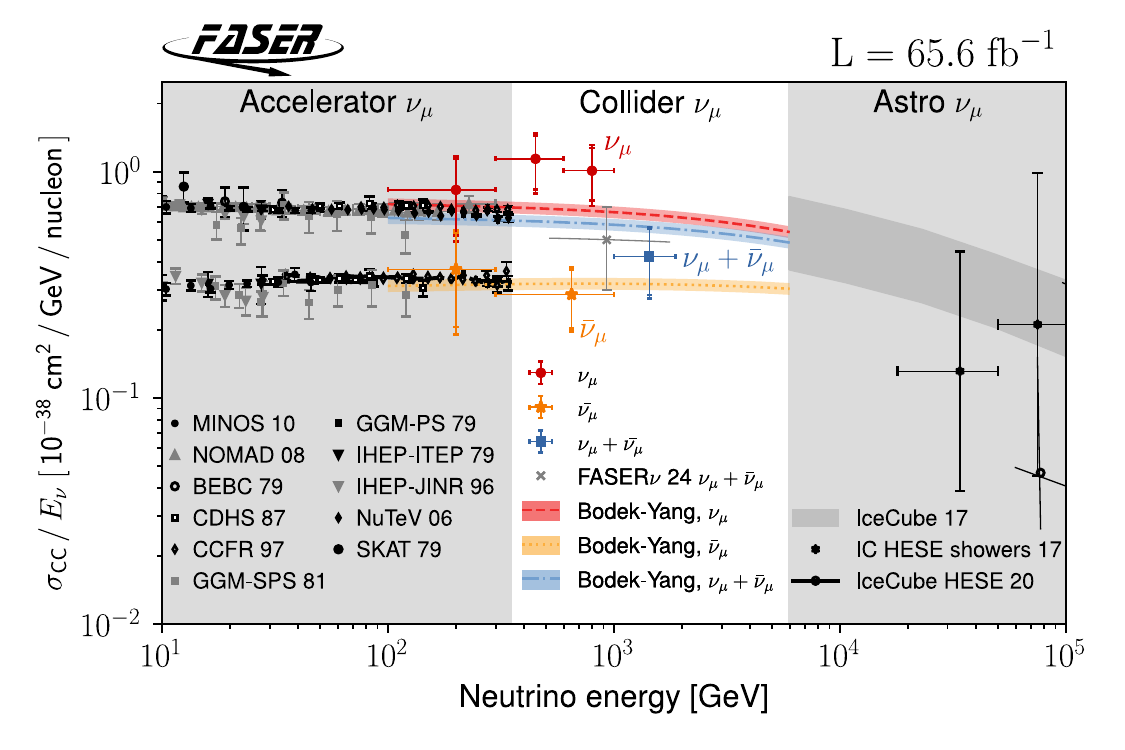} \\
  \includegraphics[width=\linewidth]{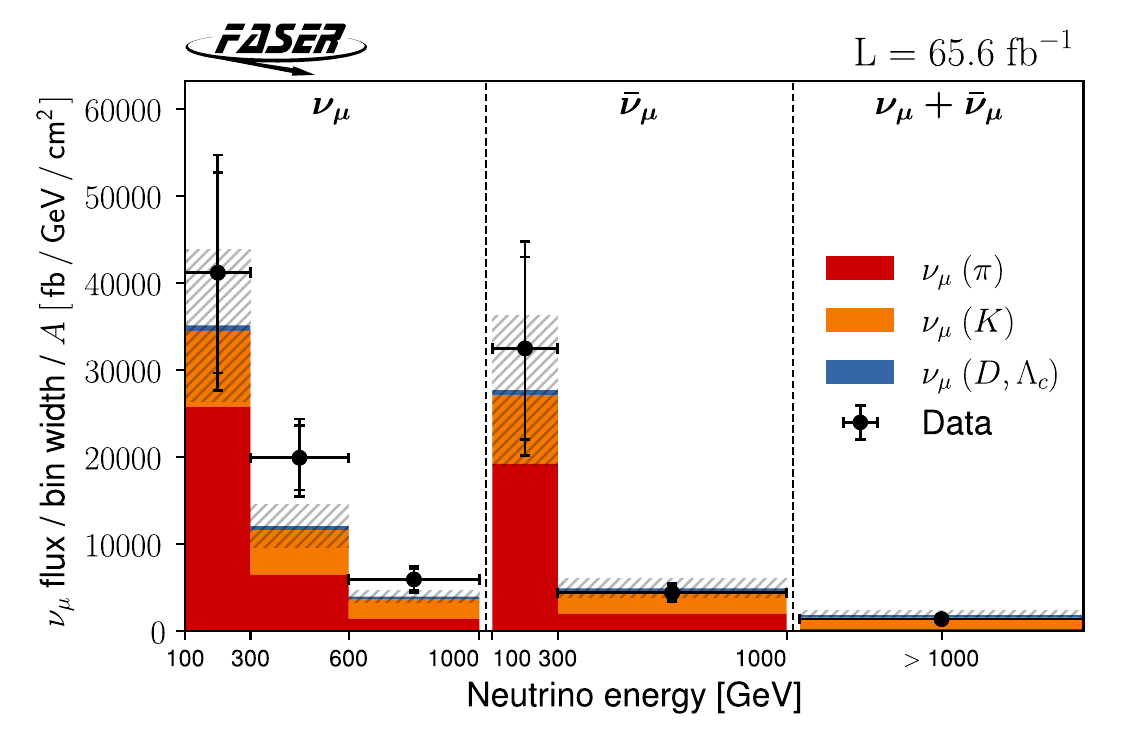}
  \vspace*{-0.15in}
  \caption{
    (\textbf{Top}) Measured CC neutrino-nucleon cross sections divided by neutrino energy as a function of neutrino energy (colored markers), compared to predictions from the Bodek-Yang model (colored lines) and previous measurements from fixed-target and astrophysical sources (gray/black markers). The horizontal error bars show the widths of the energy bins. The common high-energy bin contains neutrinos and anti-neutrinos with a momentum larger than \qty{1}{\TeV}. This corresponds to the flux-weighted average cross section. It has a width of \qty{868.2}{\GeV} which contains \SI{68.3}{\percent} of the simulated events.
    (\textbf{Bottom}) The measured neutrino flux divided by the energy and the cross sectional area of the fiducial volume (black markers) is compared to the expectation from the simulation (histograms). The inner error bars show the statistical uncertainty and the outer error bars the total uncertainty. The hatched area indicates the systematic uncertainties.
  }
  \label{fig:flux_cross-section}
\end{figure}

The measured cross sections and neutrino fluxes are shown in~\cref{fig:flux_cross-section}. The dominant uncertainties for the cross sections measurement arise from the event statistics and the sizable uncertainty of the neutrino flux. We compare the measured collider neutrino cross sections with existing measurements from fixed-target neutrino experiments~\cite{deGroot:1978feq,Colley:1979rt,Seligman:1997fe,Berge:1987zw,GargamelleNeutrinoPropane:1979kqo,GargamelleSPS:1981hpd,Mukhin:1979bd,Anikeev:1995dj,MINOS:2009ugl,NOMAD:2007krq,NuTeV:2005wsg,Baranov:1979255} at lower energies $E_\nu\lesssim \qty{360}{\GeV}$, with the measurement of the FASER$\nu$ emulsion detector~\cite{FASER:2024hoe} at \unit{\TeV} energies, as well as measurements using astrophysical neutrinos observed at IceCube~\cite{IceCube:2017roe,Bustamante:2017xuy,IceCube:2020rnc} at energies between \qty{10}{\TeV} and \qty{10}{\peta\electronvolt}. We observe good agreement between the measured and expected cross section and fluxes.
Additionally, \cref{fig:cross-section-flux-contour} shows the simultaneous fit of the cross section and flux, both with and without constraints from the expected values from simulation.

\begin{figure}[bth]
  \centering
  \includegraphics[width=\linewidth]{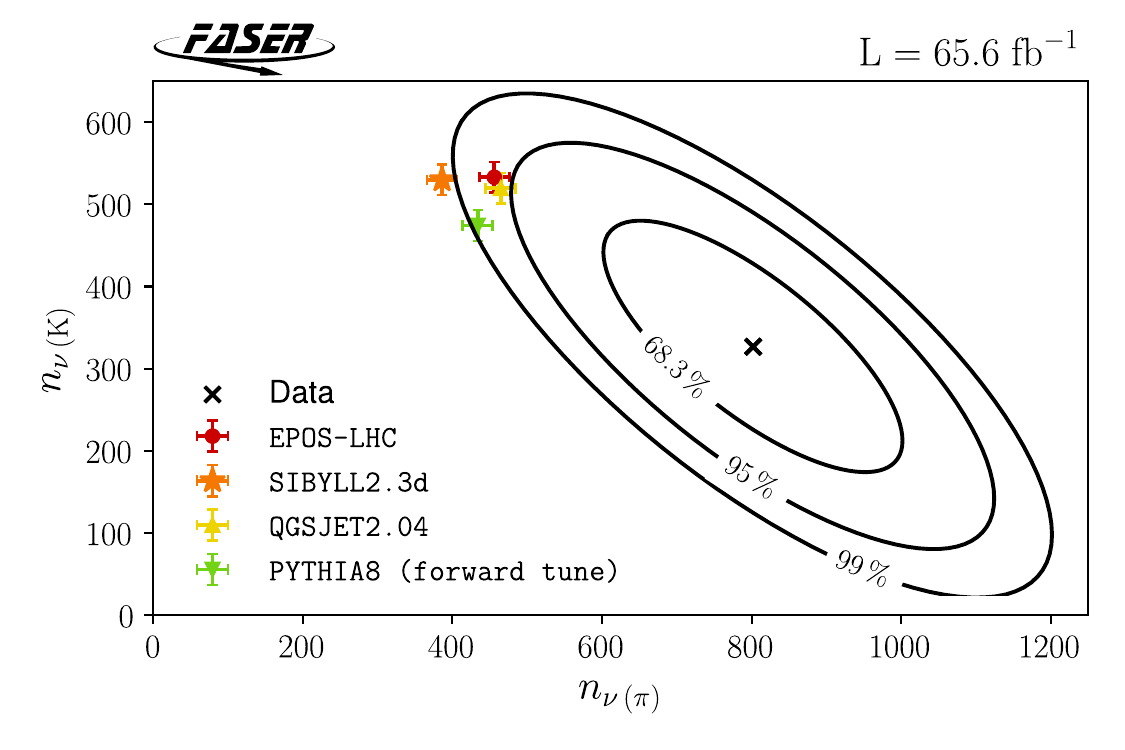}
  \vspace*{-0.15in}
  \caption{Comparison of the number of neutrino interactions from pion and kaon decays with different MC generators. The \qty{68.3}{\percent}, \qty{95}{\percent} and \qty{99}{\percent} confidence intervals, correspond to $\Delta \chi^2 = \num{2.3}$, \num{5.99} and \num{9.21}.}
  \label{fig:template-fit}
\end{figure}

The measured neutrino energy spectrum provides insight into their production mechanisms: muon neutrinos originating from pions generally have lower average energies than those from kaons or charmed mesons. This distinction is utilized to perform a $\chi^2$ fit to the unfolded number of neutrino interactions. \Cref{fig:template-fit} shows the estimated number of neutrinos from pions versus kaons, as well as the expectation from various generators. We observe a larger pion-to-kaon ratio than expected, yielding a $p$-value of \SI{5.9}{\percent} with SM expectations. This disfavors the explanations for the cosmic muon puzzle proposed in Refs.~\cite{dembinskiReportTestsMeasurements2019,albrechtMuonPuzzleCosmicray2022,anchordoquiExplanationMuonPuzzle2022,soldinUpdateCombinedAnalysis2022} that involve an enhanced forward strangeness production. More details on the template fit are given in appendix~\ref{app:pion_kaon_charm_fit}.

\papersection{Summary and Conclusions}
We present the first differential measurement of muon neutrino and anti-neutrino interactions probing neutrinos in the \unit{\TeV} energy range. The result was obtained analyzing \qty{65.6 +- 1.4}{\ifb} of LHC $pp$ collision data, recorded during 2022 and 2023. Neutrino candidates are selected using the active electronic components of the FASER detector. We observe in total \num{338.1 +- 21.0} neutrino events from CC \Pnum and \APnum interactions in the fiducial volume, which agrees with the expectation within the uncertainties.

The measured muon momentum, $q/p^{\prime}_{\Pmu}$, is unfolded to the neutrino energy in six bins covering energies from \qty{100}{\GeV} to the \unit{\TeV} range. With this and the predicted neutrino and anti-neutrino flux the energy-dependent neutrino nucleon cross section is determined. Conversely, the predicted neutrino- and anti-neutrino-nucleon cross section is used to determine the differential muon neutrino flux. The contribution of neutrinos from pion and kaon decays is estimated using a template fit, finding a larger flux fraction for neutrinos from pions than kaons. These results together with the recent cross section measurement using the FASER$\nu$ emulsion detector~\cite{FASER:2024hoe} represent the first studies of the neutrino-nucleon cross section and neutrino flux using collider neutrinos probing the \unit{\TeV} energy range, with wide ranging implications~\cite{Worcester:2023njy}. The presented results and the full experimental covariance matrix are available on \texttt{HEPData} \cite{fasercollaborationFirstMeasurementMuon2025}.

\papersection{Acknowledgments}
We thank CERN for the very successful operation of the LHC during 2022 and 2023. We thank the technical and administrative staff members at all FASER institutions for their contributions to the success of the FASER project. We thank the ATLAS Collaboration for providing us with accurate luminosity estimates for the used Run 3 LHC collision data. FASER gratefully acknowledges the donation of spare ATLAS SCT modules and spare LHCb calorimeter modules, without which the experiment would not have been possible. We also acknowledge the use of the ATLAS Collaboration software, Athena, on which FASER's offline software system is based~\cite{ATL-PHYS-PUB-2009-011} and the use of the ACTS tracking software framework~\cite{Ai:2021ghi}. Finally we thank the CERN STI group for providing detailed FLUKA simulations of the muon fluence along the LOS, which have been used in this analysis. This work was supported in part by Heising-Simons Foundation Grant Nos.~2018-1135, 2019-1179, and 2020-1840, Simons Foundation Grant No.~623683, U.S.~National Science Foundation Grant Nos.~PHY-2111427, PHY-2110929, and PHY-2110648, JSPS KAKENHI Grant Nos.~19H01909, 22H01233, 20K23373, 23H00103, 20H01919, and 21H00082, the joint research program of the Institute of Materials and Systems for Sustainability, ERC Consolidator Grant No.~101002690, Horizon 2020 Framework Programme Grant No.~MUCCA CHIST-ERA-19-XAI-00, BMBF Grant No.~05H20PDRC1, DFG EXC 2121 Quantum Universe Grant No.~390833306, Royal Society Grant No.~URF\textbackslash R1\textbackslash 201519, U.K. Science and Technology Funding Councils Grant No.~ST/T505870/1, the U.K. Science and Technology Funding Council, the National Natural Science Foundation of China, Tsinghua University Initiative Scientific Research Program, and the Swiss National Science Foundation.

\bibliographystyle{utphys}
\bibliography{references.bib}

\clearpage
\newpage

\onecolumngrid

\section*{Appendix}

\begin{appendix}

\section{Geometric Muon Background} \label{app:geo_bkg}
\begin{figure}[ht]
  \centering
  \includegraphics[width=0.45\linewidth]{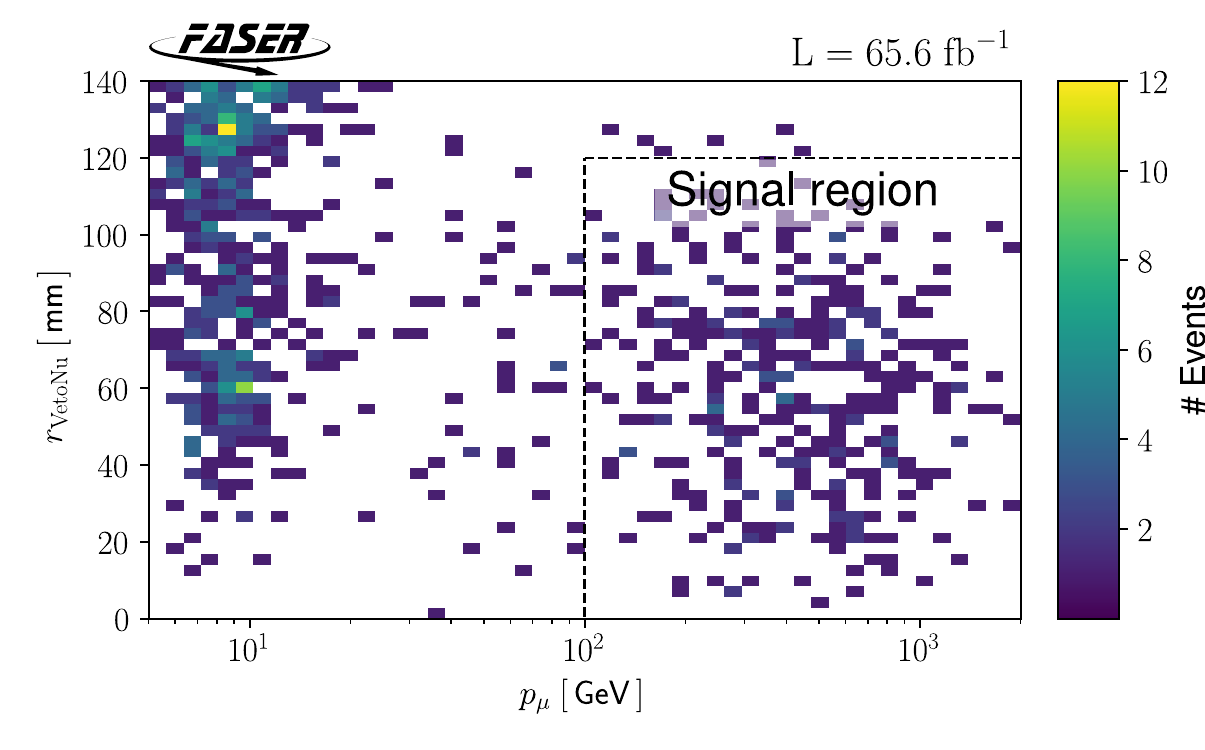}
  \includegraphics[width=0.42\textwidth]{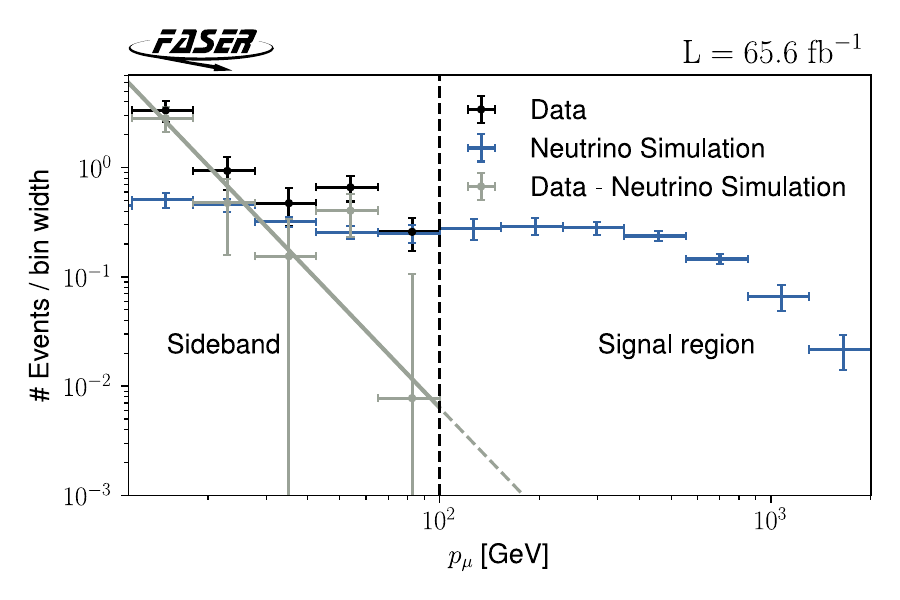}
  \vspace*{-0.15in}
  \caption{
    (\textbf{Left}) The distribution of the extrapolated, transverse track position at the VetoNu scintillator station, $r_{\rm VetoNu}$, and momentum, $p_{\Pmu}$. The black box shows the signal region.
    (\textbf{Right}) The fit to the geometric muon background in the region $p_{\mu} < \qty{100}{\GeV}$ is shown for negatively charged tracks. The simulated neutrino prediction (blue) is subtracted from the data (black), and the result (gray) is fitted and extrapolated to $p_{\mu} > \qty{100}{\GeV}$.
  }
  \label{fig:geo_bkg}
\end{figure}

The distribution of the transverse distance from the detector axis of the track when extrapolated to the VetoNu scintillator, $r_{\rm VetoNu}$, and the track momentum, $p_{\Pmu}$, in \cref{fig:geo_bkg} (left) shows a clear separation between signal events (black dashed box), and background events with momenta $\lesssim \qty{20}{\GeV}$. Events with a momentum between \qty{20}{\GeV} and \qty{100}{\GeV} are consistent with neutrino simulation.

\Cref{fig:geo_bkg} (right) shows the extrapolation of the geometric muon background from the sideband into the signal region. The neutrino expectation is subtracted from the data and fitted with an exponential function. Systematic variations in the extrapolation fit were also tested, but as the resulting change in the background predictions were within the data statistical uncertainties no additional uncertainty was applied.

\FloatBarrier

\section{Response Matrix and Efficiency}\label{app:response_matrix}
\begin{figure*}[ht]
  \centering
  \includegraphics[width=0.5\textwidth]{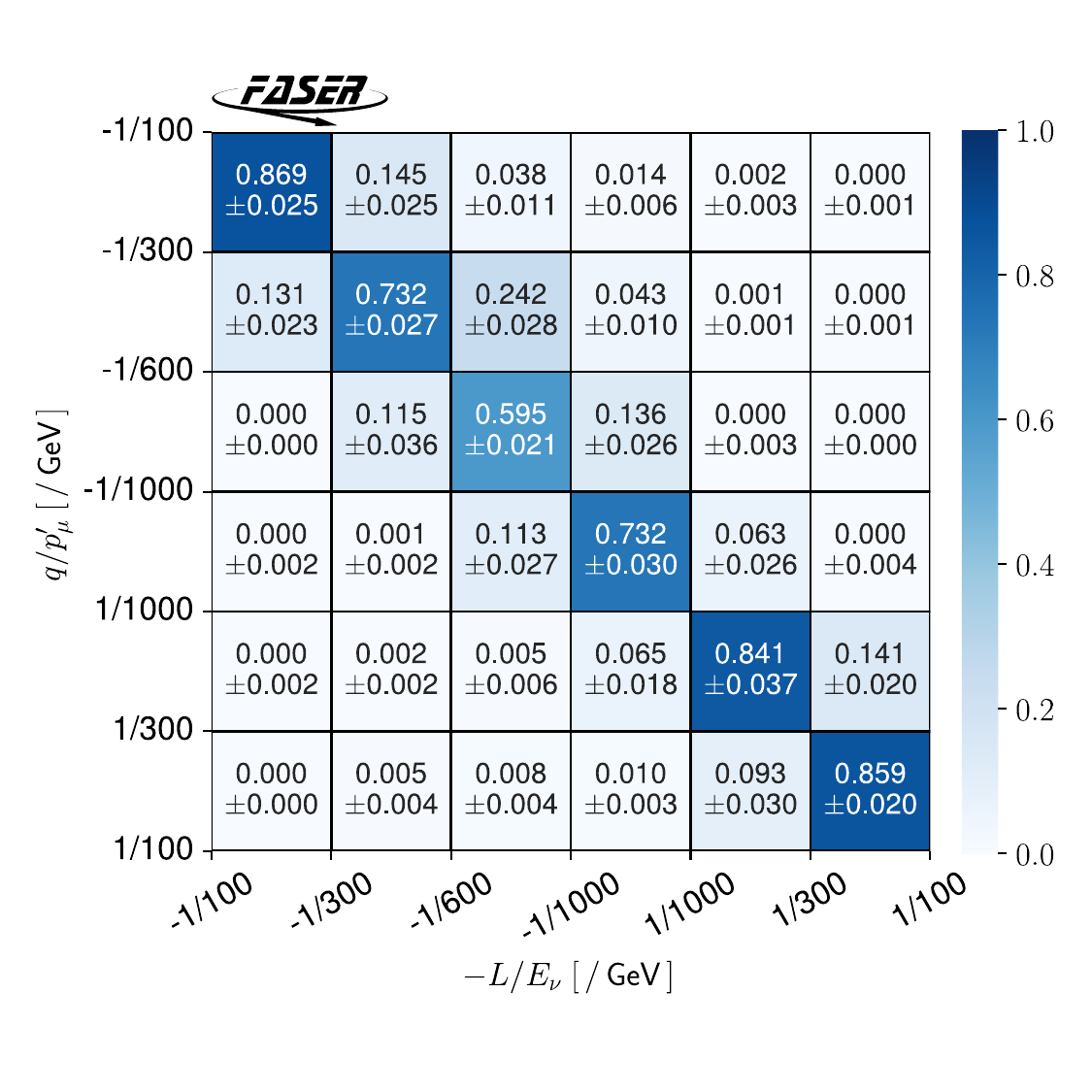}
  \vspace*{-0.15in}
  \caption{The response matrix relating the reconstructed, calibrated muon momentum $q/p^{\prime}_{\Pmu}$ to the neutrino energy $-L/E_{\Pnu}$, expressed as a conditional probability. The off-diagonal elements are dominated by the fluctuations in the fraction of the neutrino energy that the muon takes in the interaction. The uncertainties reflect the quadratic sum of statistical and systematic uncertainties and are dominated by the modeling of the neutrino interactions (these uncertainties are about \num{2.5} times larger than detector related uncertainties, such as alignment).}
  \label{fig:mig_matrix}
\end{figure*}

\Cref{fig:mig_matrix} shows the response matrix relating the calibrated, reconstructed muon momentum $q/p^{\prime}_{\Pmu}$ to the neutrino energy $-L/E_{\Pnu}$, expressed as a conditional probability. Using the calibrated muon momentum, $p^{\prime}_{\Pmu} = p_{\Pmu} / 0.8$, results in a more diagonal response matrix, and therefore smaller uncertainties on the unfolded number of neutrino interactions.

\Cref{tab:reconstruction-efficiency} shows the detector acceptance and reconstruction efficiency.
The acceptance is the fraction of neutrino interactions in the fiducial volume where the produced muon has a momentum larger \qty{100}{\GeV} and goes through the active transverse area of the whole detector at truth level. The transverse momentum of the muon produced in the neutrino interaction is particularly relevant for low-energy neutrinos, resulting in smaller acceptances in the corresponding bins. In addition, negatively charged muons typically get a smaller fraction of the neutrino energy, resulting in an asymmetry between neutrinos and anti-neutrinos. The reconstruction efficiency is the fraction of these events which pass the event selection. The inefficiency is dominated by the angular and the VetoNu charge cut. The efficiency used in \cref{eq:unfolding} is the product of acceptance and reconstruction efficiency.

\begin{table}[ht]
  \centering
  \caption{Acceptance, reconstruction efficiency, and total efficiency in percent in bins of the neutrino energy, $-L/E_{\Pnu}$. The uncertainties describe the modeling of the neutrino flux and interactions, the change of the LOS, and the detector response.}
  \label{tab:reconstruction-efficiency}
  \begin{ruledtabular}
    \begin{tabular}{lS[table-format=2.1(1.1)]S[table-format=2.1(1.1)]S[table-format=2.1(1.1)]S[table-format=2.1(1.1)]S[table-format=2.1(1.1)]S[table-format=2.1(1.1)]S[table-format=2.1(1.1)]}
      $-L/E_{\Pnu} \ \left[\si{\per\GeV}\right]$ & {{$\left[\frac{-1}{100}, \frac{-1}{300}\right]$}} & {{$\left[\frac{-1}{300}, \frac{-1}{600}\right]$}} & {{$\left[\frac{-1}{600}, \frac{-1}{1000}\right]$}} & {{$\left[\frac{-1}{1000}, \frac{1}{1000}\right]$}} & {{$\left[\frac{1}{1000}, \frac{1}{300}\right]$}} & {{$\left[\frac{1}{300}, \frac{1}{100}\right]$}} & {{ Total }} \\
      \midrule
      Acceptance $\alpha$ [\si{\percent}]                                       & 16.7+-0.6 & 30.0+-1.2 & 38.8+-2.6 & 46.6+-3.0 & 47.5+-1.9 & 27.8+-2.5 & 35.3+-2.0 \\
      Reco. efficiency $\epsilon_{\rm reco.}$ [\si{\percent}]                   & 80.8+-7.2 & 79.0+-4.9 & 79.3+-5.0 & 78.8+-5.5 & 84.4+-4.7 & 84.5+-6.9 & 80.3+-4.5 \\
      \midrule
      Efficiency $\epsilon = \alpha \cdot \epsilon_{\rm reco}$  [\si{\percent}] & 13.5+-1.2 & 23.7+-1.7 & 30.8+-2.6 & 36.8+-3.0 & 40.1+-3.1 & 23.5+-2.4 & 28.3+-1.9\\
    \end{tabular}
  \end{ruledtabular}
\end{table}

\FloatBarrier

\section{Flux and neutrino interaction modeling uncertainties} \label{app:modeling_unc}
The primary source of the neutrino flux uncertainty originates from the modeling of the forward hadron rate. To estimate this uncertainty for light hadron production, we follow Ref.~\cite{fasercollaborationNeutrinoRatePredictions2024} and consider the range of predictions from \texttt{EPOS-LHC}, \texttt{Sibyll~2.3d}~\cite{Riehn:2019jet}, \texttt{QGSJET~2.04}~\cite{Ostapchenko:2010vb}, and the forward physics tune of \texttt{PYTHIA}~\cite{Fieg:2023kld}. For heavy hadron production we follow Ref.~\cite{buonocorePredictionsNeutrinosNew2024} and vary the factorization and renormalization scales by a factor of 2 around their nominal value.

The neutrino interactions with the detector material are simulated with \texttt{GENIE~3.04.0}~\cite{Andreopoulos:2009rq,Andreopoulos:2015wxa,GENIE:2021zuu}. To estimate the uncertainty from the modeling of final state interactions and intranuclear hadron transport, two different tunes are applied using the default \texttt{INTRANUKE hA} effective intranuclear rescattering model and the \texttt{INCL++} implementation of the Li\`{e}ge intranuclear model~\cite{mancusi:2015}. The passage of the particles produced in neutrino interactions through the detector is simulated with \texttt{GEANT4}. To estimate the uncertainty from hadron transport~\cite{mendoza:2014newSE,mendoza:2018}, we study the \texttt{FTFP\_INCLXX}, \texttt{FTF\_BIC}, and \texttt{FTFP\_BERT\_HP} physics lists together with the \texttt{G4NDL}, \texttt{JEFF-3.3}~\cite{plompen:2020} and \texttt{ENDF/B-VIII.0}~\cite{brown:2018} neutron data libraries.

\section{Comparison of signal and background distributions} \label{app:sig_bkg_comp}
\begin{figure*}[ht]
  \includegraphics[width=0.33\textwidth]{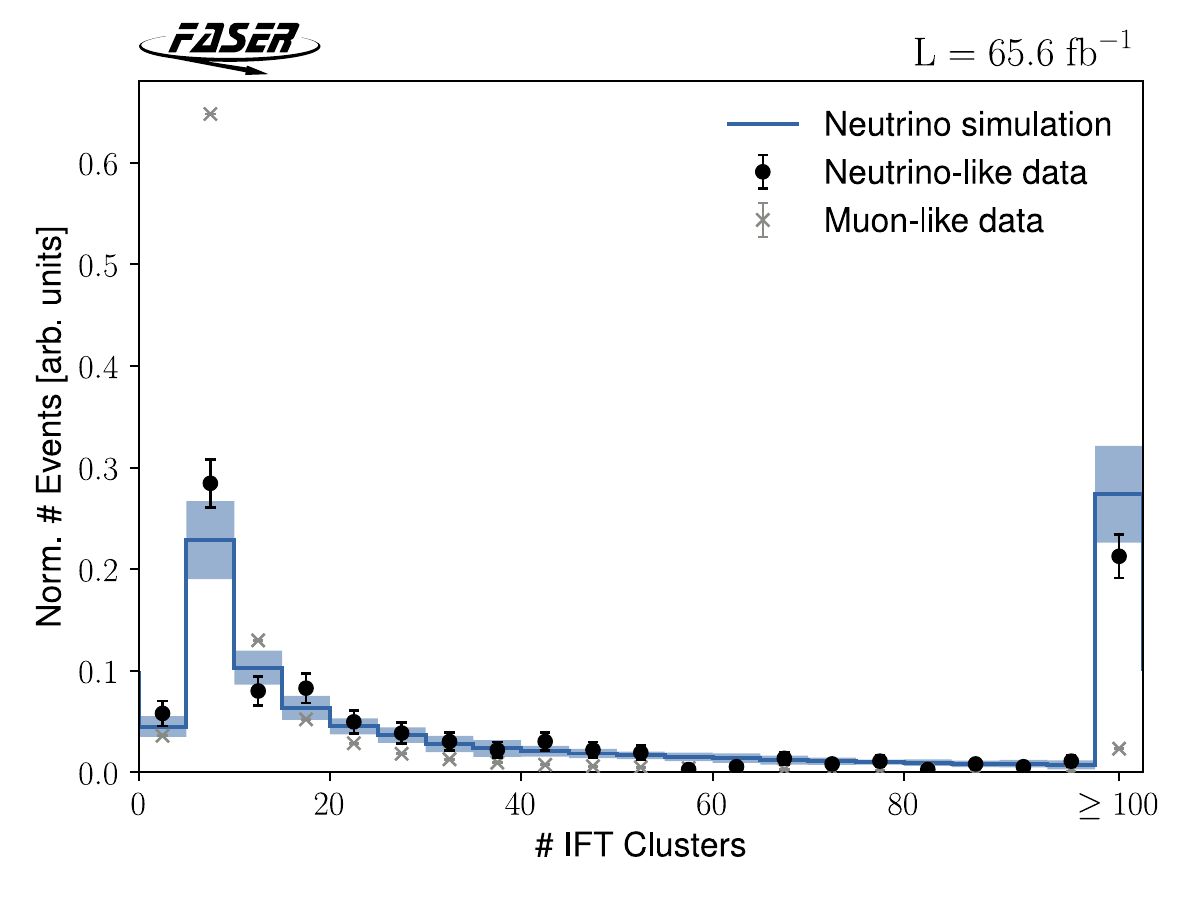}%
  \includegraphics[width=0.33\textwidth]{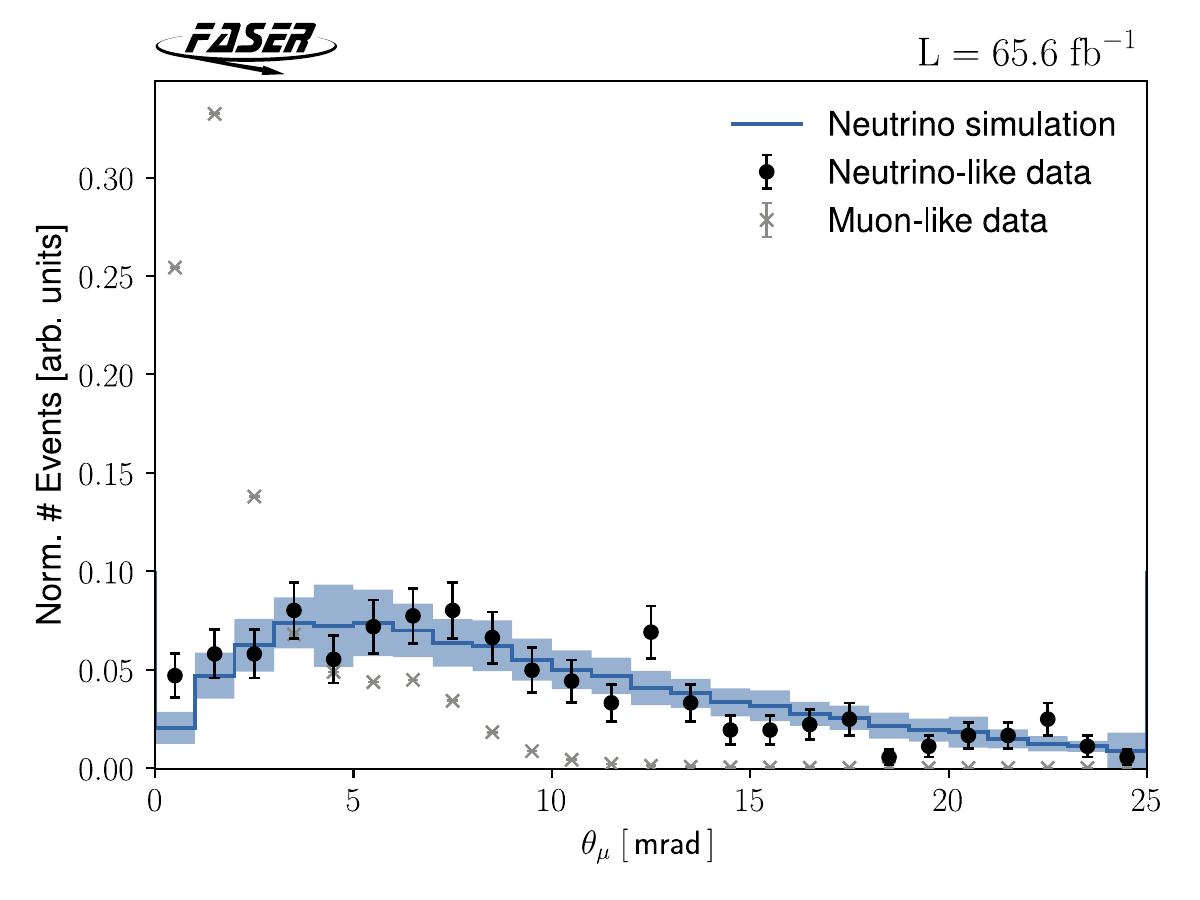}%
  \includegraphics[width=0.33\textwidth]{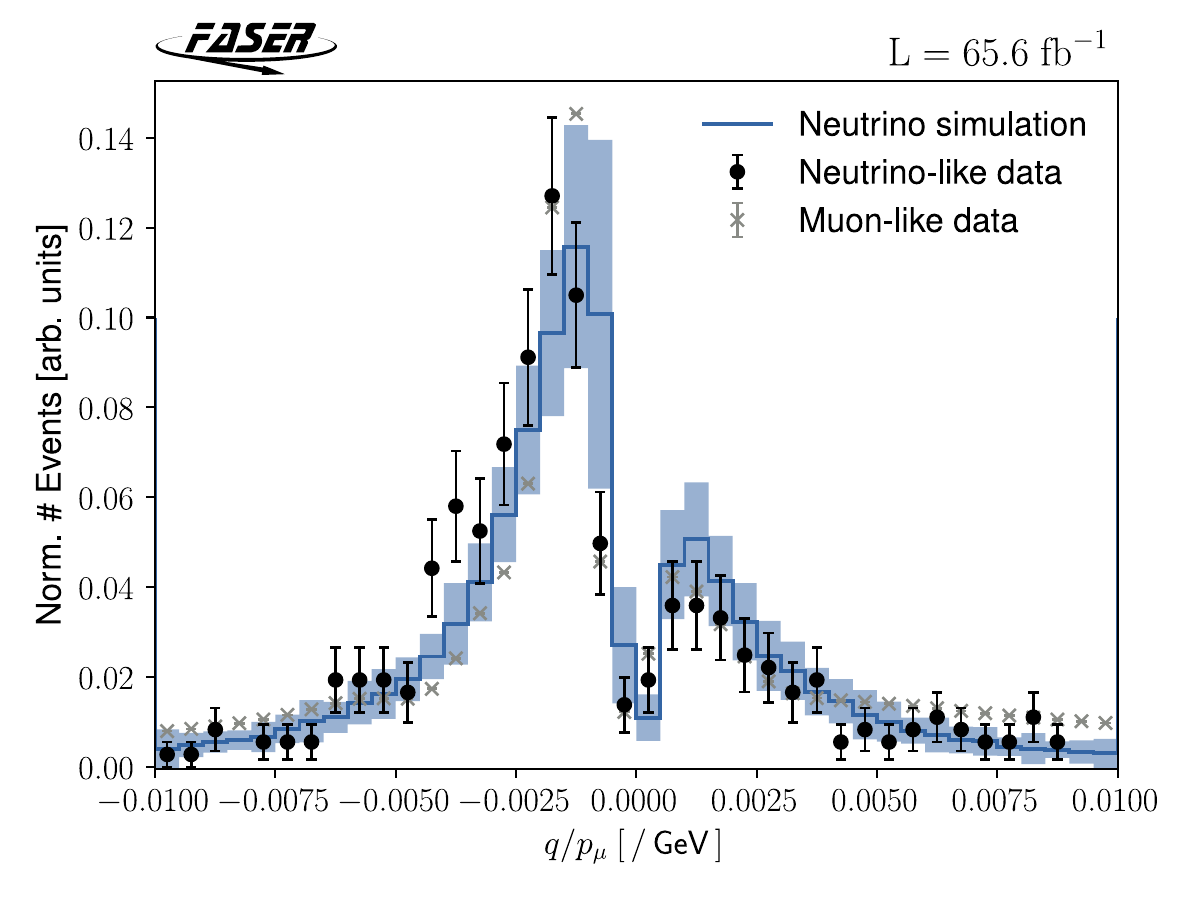}
  \vspace*{-0.15in}
  \caption{Number of clusters in the IFT (\textbf{left}), track polar angle $\theta_{\Pmu}$ (\textbf{center}) at the first station of the tracking spectrometer, and the ratio of the muon unit charge and momentum $q/p_{\Pmu}$ (\textbf{right}) for signal events (black markers), muon-like events (gray markers), and simulation (blue histogram). The muon-like data pass the same event-selection as signal events, but have a reduced charge larger than \qty{30}{\pico\coulomb} in both layers of the VetoNu scintillator station. The uncertainty of the simulated events describes the modeling of the neutrino flux and neutrino interactions, the change of the LOS, the detector response, and the luminosity and mass uncertainties.}
  \label{fig:muon-distribution}
\end{figure*}

We compare the distribution of the signal events, with simulation and background-like events. In the following signal refers to muons produced in a neutrino interaction in the tungsten target, and background muons refers to muons from the decay of particles produced at the ATLAS IP. For the background, we select events which pass the event selection but have a reduced charge larger than \qty{30}{\pico\coulomb} in both layers of the VetoNu scintillator: (i) Background events have usually only a single muon traversing the IFT, which creates one hit in each wafer, resulting in a total of six hits. In contrast, the neutrino interactions can produce an electromagnetic and hadronic shower which may hit the IFT, resulting in a large number of hits. (ii) Muons from the decay of particles produced at the ATLAS IP are only slightly deflected by the LHC magnets (otherwise they miss the FASER detector), resulting in small angles with respect to the LOS. In contrast, muons produced in a neutrino interaction in the tungsten target can have much larger angles. (iii) Muons from background and signal events have slightly different $q/p_{\Pmu}$ distributions. As shown in \cref{fig:muon-distribution}, for these distributions signal events agree with simulation, but are very distinct from background-like events.

\FloatBarrier

\section{Fit of neutrinos from pions and kaons}\label{app:pion_kaon_charm_fit}
The unfolded number of neutrino interactions is analyzed to determine the fractions of neutrinos originating from pion and kaon decays. The expected number of neutrino interactions in bin \(i\) of the unfolded neutrino energy can be calculated as \( \nu_{\Pnu}^{i} = \sum_{k} f_{ik} \eta_{k} \, ,\) where \(\eta_k\) denotes the total number of neutrino interactions from production mode \(k\), and \(f_{ik}\) represents the fraction of these neutrino interactions in bin \(i\), obtained from simulation. Since the expected contribution from charmed hadrons is small and the shapes of charm and kaon contributions are similar, we fix it to the expected value within uncertainties. To determine the number of neutrino interactions, we perform a \(\chi^2\) fit with
\begin{equation}\label{eq:chi2-fit}
  \chi^2 = \left( \vec{\nu} - \vec{n} \right)^T {\rm Cov}^{-1} \left( \vec{\nu} - \vec{n} \right) + \mathcal{G} \, ,
\end{equation}
where \(\vec{n}\) and \({\rm Cov}\) represent the number and covariance matrix of neutrino interactions obtained from the likelihood fit in \cref{eq:likelihood}. Systematic uncertainties on the fractions \(f_{ik}\) are implemented with nuisance parameters and are constrained with Gaussian priors \(\mathcal{G}\), but correlations between different templates are not taken into account. They are dominated by the shape uncertainties from different generators.

\FloatBarrier

\end{appendix}

\clearpage

\section*{Supplemental Material}
\setcounter{section}{0}

\section{Reduced vetoNu charge} \label{app:reduced-vetonu-charge}
\begin{figure}[h]
  \centering
  \includegraphics[width=0.6\textwidth]{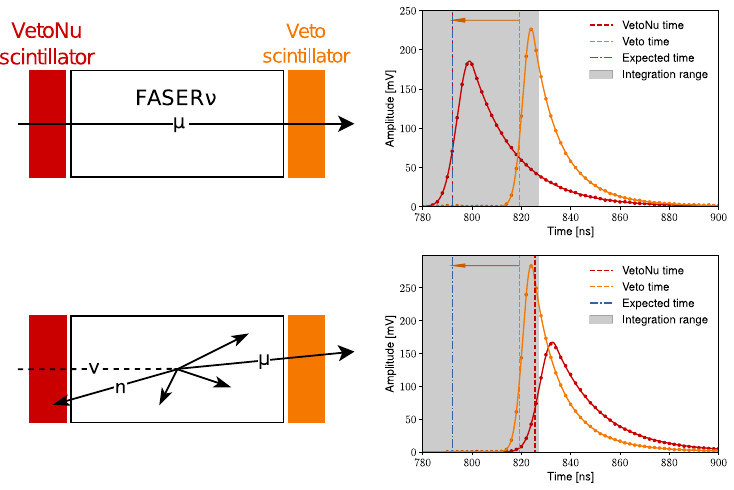}
  \vspace*{-0.15in}
  \caption{
    (\textbf{Left}) Depiction of a muon (top) and a neutrino (bottom) traversing the FASER$\nu$ detector and depositing charge in the vetoNu (red) and veto (orange) scintillator stations.
    (\textbf{Right}) Waveforms of the vetoNu and veto charge. The red and orange lines show the time where the rising edge of the amplitude is \qty{40}{\percent} of the maximum value. The blue line shows the expected time, which is calculated from the measured local veto time and the average time-of-flight of muons. To cover also the tail of the waveforms a constant \qty{35}{\ns} offset is added to the expected time, which is indicated by the gray area. The reduced charge is defined as the measured charge in each of the two vetoNu scintillators integrated over this time window.
  }
  \label{fig:reduced-charge-depiction}
\end{figure}

The left panel in \cref{fig:reduced-charge-depiction} illustrates how a muon and neutrino deposit charge in the vetoNu and veto scintillators. The right panel shows the simulated waveforms in these scintillator stations for a typical muon and neutrino event and explains the calculation of the reduced vetoNu charge.

\FloatBarrier

\section{Simulated neutrino flux and effective cross section}
\Cref{tab:neutrino-flux-cross-section} shows the simulated neutrino flux and effective cross section in bins of the neutrino energy. The uncertainty of the simulated flux is dominated by the modeling from different generators. The effective cross section is the flux-weighted average and depends on the modeling within each energy bin. A systematic uncertainty is applied to account for the difference between the used generators. Additionally, a \qty{6}{\percent} uncertainty is applied to account for the difference between the Bodek-Yang model~\cite{Bodek:2002vp,Bodek:2004pc,Bodek:2010km} employed in \texttt{GENIE} and more recent cross section models~\cite{Candido:2023utz,Jeong:2023hwe}.

\begin{table}[h]
  \caption{Simulated neutrino flux and effective cross section}
  \label{tab:neutrino-flux-cross-section}
  \centering
  \begin{ruledtabular}
    \begin{tabular}{lS[table-format=3.1(1.1)]S[table-format=3.1(1.1)]S[table-format=3.1(1.1)]S[table-format=3.1(1.1)]S[table-format=3.1(1.1)]S[table-format=2.1(1.1)]}
      $-L/E_{\Pnu} \ \left[\si{\per\GeV}\right]$ & {{$\left[\frac{-1}{100}, \frac{-1}{300}\right]$}} & {{$\left[\frac{-1}{300}, \frac{-1}{600}\right]$}} & {{$\left[\frac{-1}{600}, \frac{-1}{1000}\right]$}} & {{$\left[\frac{-1}{1000}, \frac{1}{1000}\right]$}} & {{$\left[\frac{1}{1000}, \frac{1}{300}\right]$}} & {{$\left[\frac{1}{300}, \frac{1}{100}\right]$}} \\
      \midrule
      $\phi_{\rm Sim.}$ [$10^{6}$\si{\fb\per\cm\squared}]         & 7.0+-1.8   & 3.6+-0.7    & 1.6+-0.3    & 1.7+-0.5    & 3.5+-0.8    & 5.5+-1.7  \\
      $\sigma_{\rm sim.}$ [$10^{-38}$ \si{\cm\squared} / nucleon] & 128.5+-7.9 & 292.2+-17.6 & 514.6+-31.3 & 799.3+-68.1 & 166.3+-10.5 & 56.8+-3.4 \\
    \end{tabular}
  \end{ruledtabular}
\end{table}

\FloatBarrier

\clearpage

\section{Comparison of \texttt{EPOS-LHC}+\texttt{POWHEG}+\texttt{PYTHIA8} and \texttt{DPMJET}} \label{app:dpmjet}
\begin{figure}[h]
  \centering
  \includegraphics[width=0.5\textwidth]{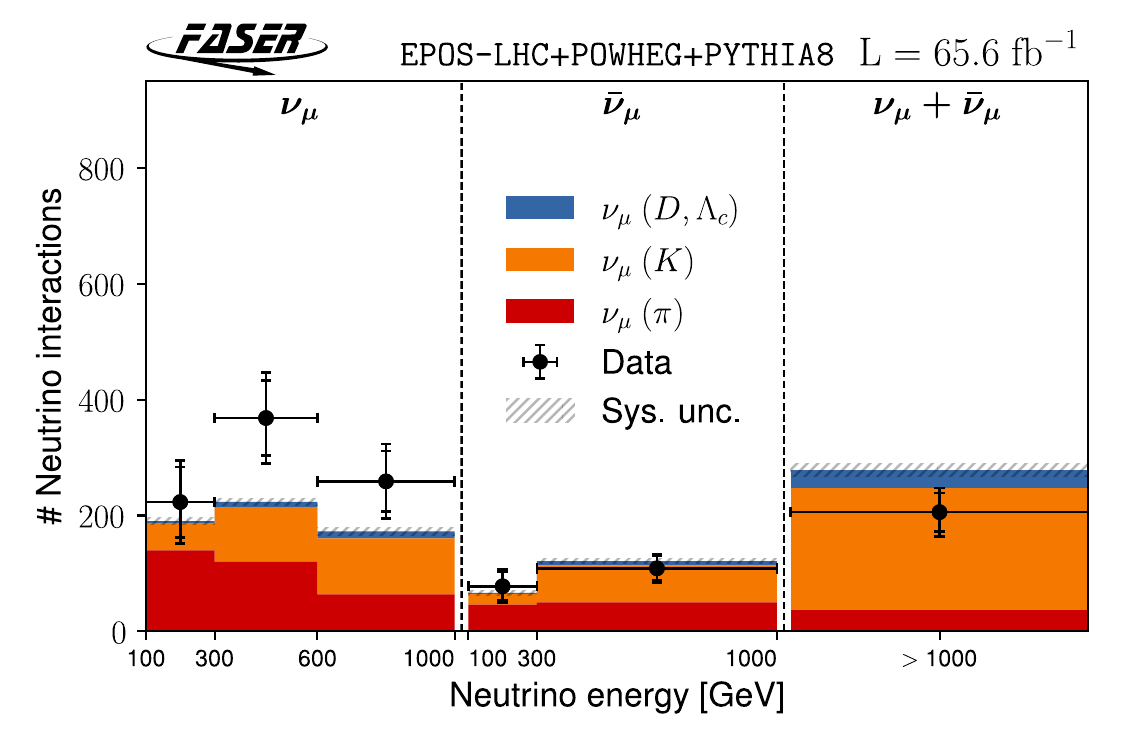}%
  \includegraphics[width=0.5\textwidth]{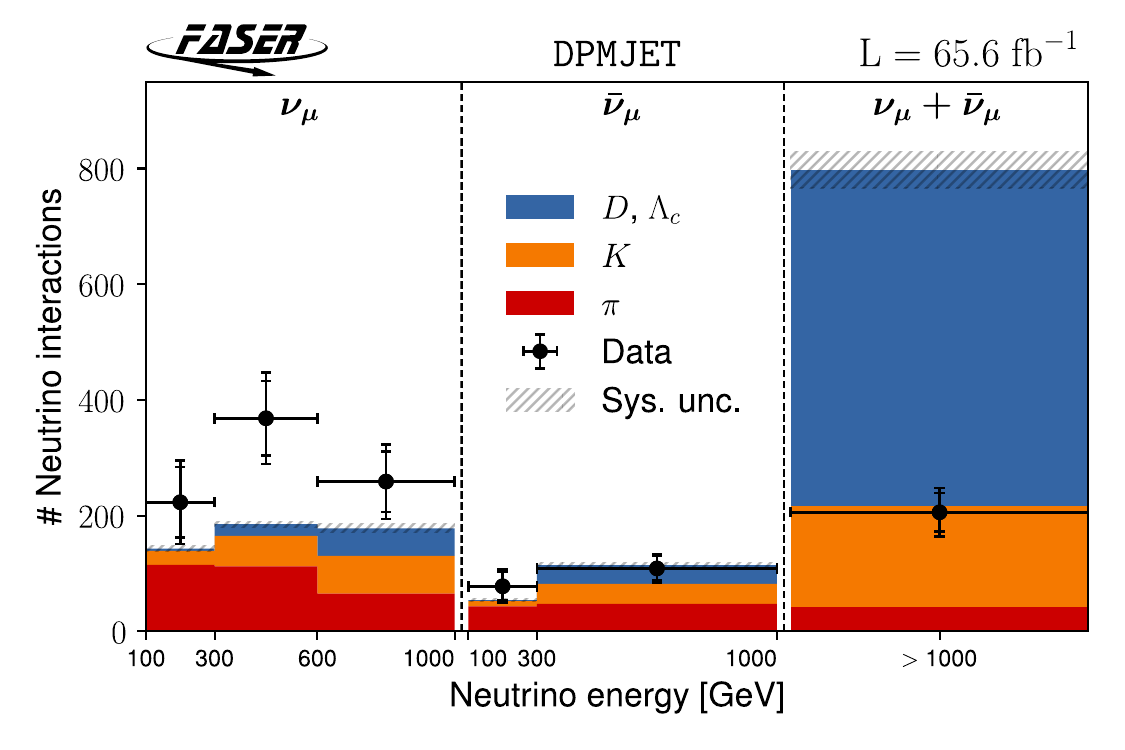}
  \vspace*{-0.15in}
  \caption{The observed number of neutrino interactions (black markers) is compared to the expectation from \texttt{EPOS-LHC} for neutrinos from the decay of pions and kaons and \texttt{POWHEG} and \texttt{PYTHIA~8.3} for the decay of charmed hadrons (\textbf{left}) and \texttt{DPMJET 3.2019.1} (\textbf{right}). The inner error bars show the statistical uncertainty and the outer error bars the total uncertainty. The hatched area in both plots indicates the systematic uncertainty without the neutrino flux modeling uncertainty (which was previously calculated from the difference of various generators).}
  \label{fig:dpmjet-num-nu-interactions}
\end{figure}

\Cref{fig:dpmjet-num-nu-interactions} compares the unfolded number of neutrino interactions with the expectation from \texttt{EPOS-LHC}, \texttt{POWHEG}, and \texttt{PYTHIA8} and with the expectation from \texttt{DPMJET 3.2019.1}.

\FloatBarrier

\section{Pre- and Post-fit distributions for the pion/kaon fit}
\begin{figure}[h]
  \centering
  \includegraphics[width=0.5\textwidth]{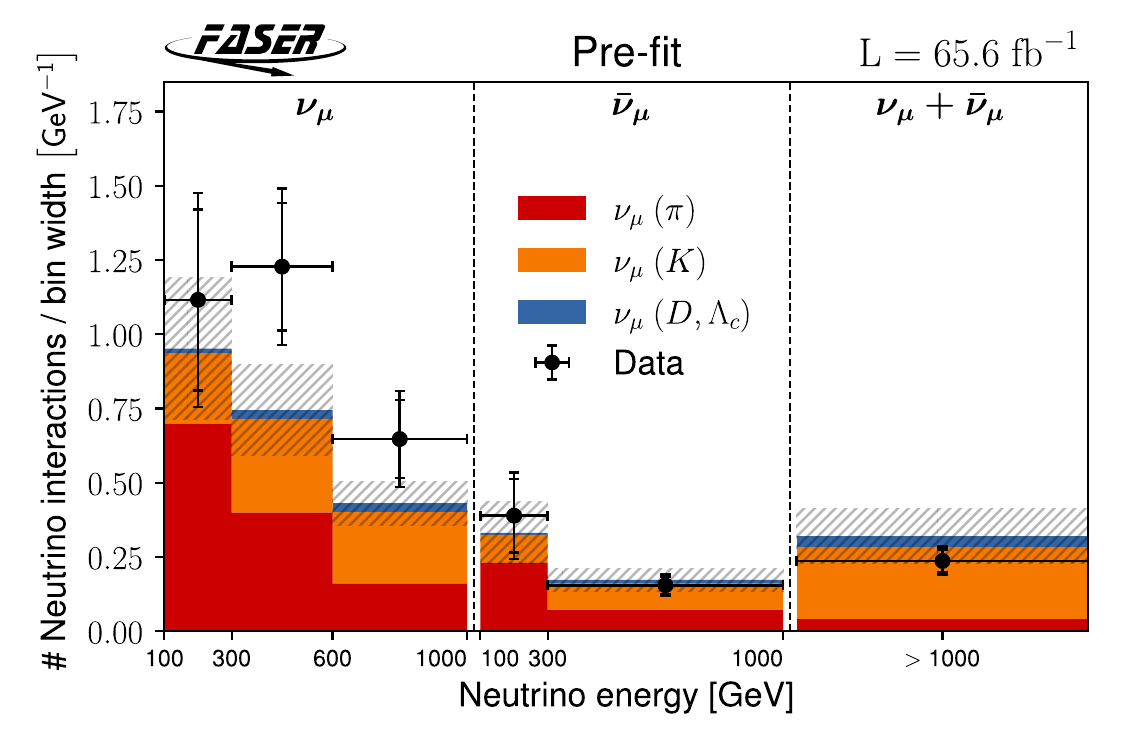}%
  \includegraphics[width=0.5\textwidth]{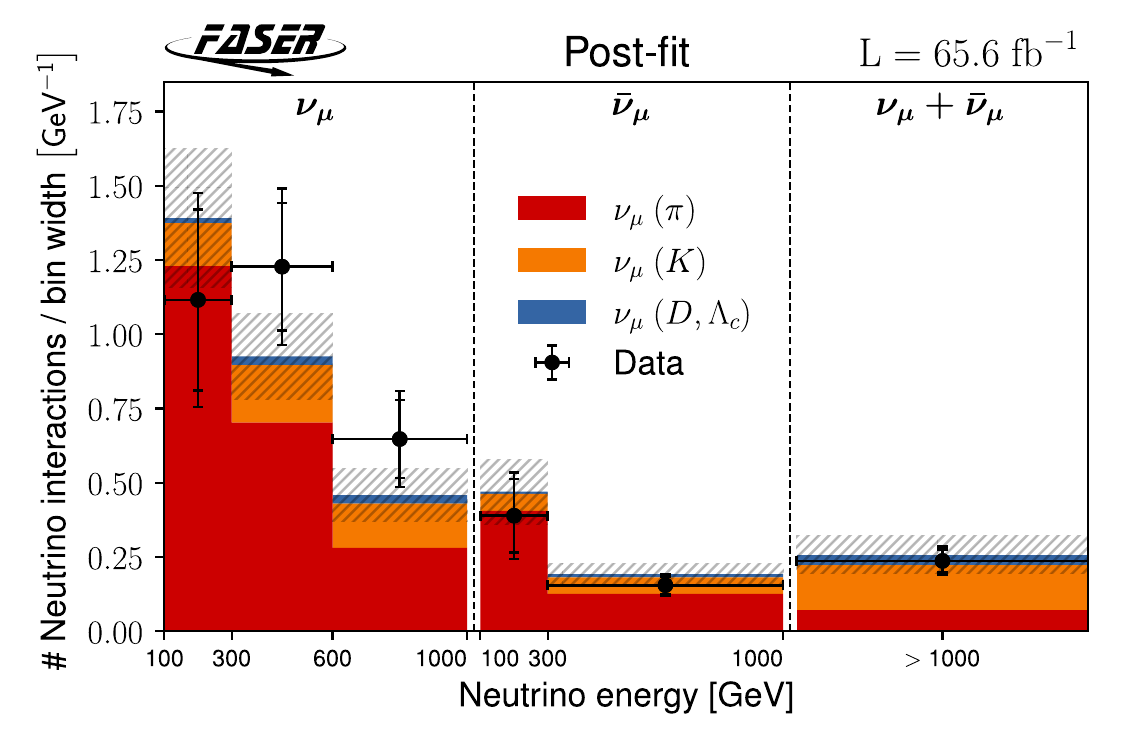}
  \vspace*{-0.15in}
  \caption{The observed number of neutrino interactions (black markers) is compared to the pre-fit expectation with a total of \num{456.0+-70.6} neutrinos from pion decays and \num{532.9+-59.3} neutrinos from kaon decays (\textbf{left}) and the post-fit expectation with a total of \num{802.0+-131.8} neutrinos from pion decays and \num{326.6+-100.9} neutrinos from kaon decays (\textbf{right}).}
  \label{fig:prefit-postfit-comparison-num-nu-interactions}
\end{figure}

\Cref{fig:prefit-postfit-comparison-num-nu-interactions} compares the unfolded number of neutrino interactions with the pre- and post-fit expectations.

\FloatBarrier

\clearpage

\section{Simultaneous fit}
\begin{figure*}[ht]
  \includegraphics[width=0.32\textwidth]{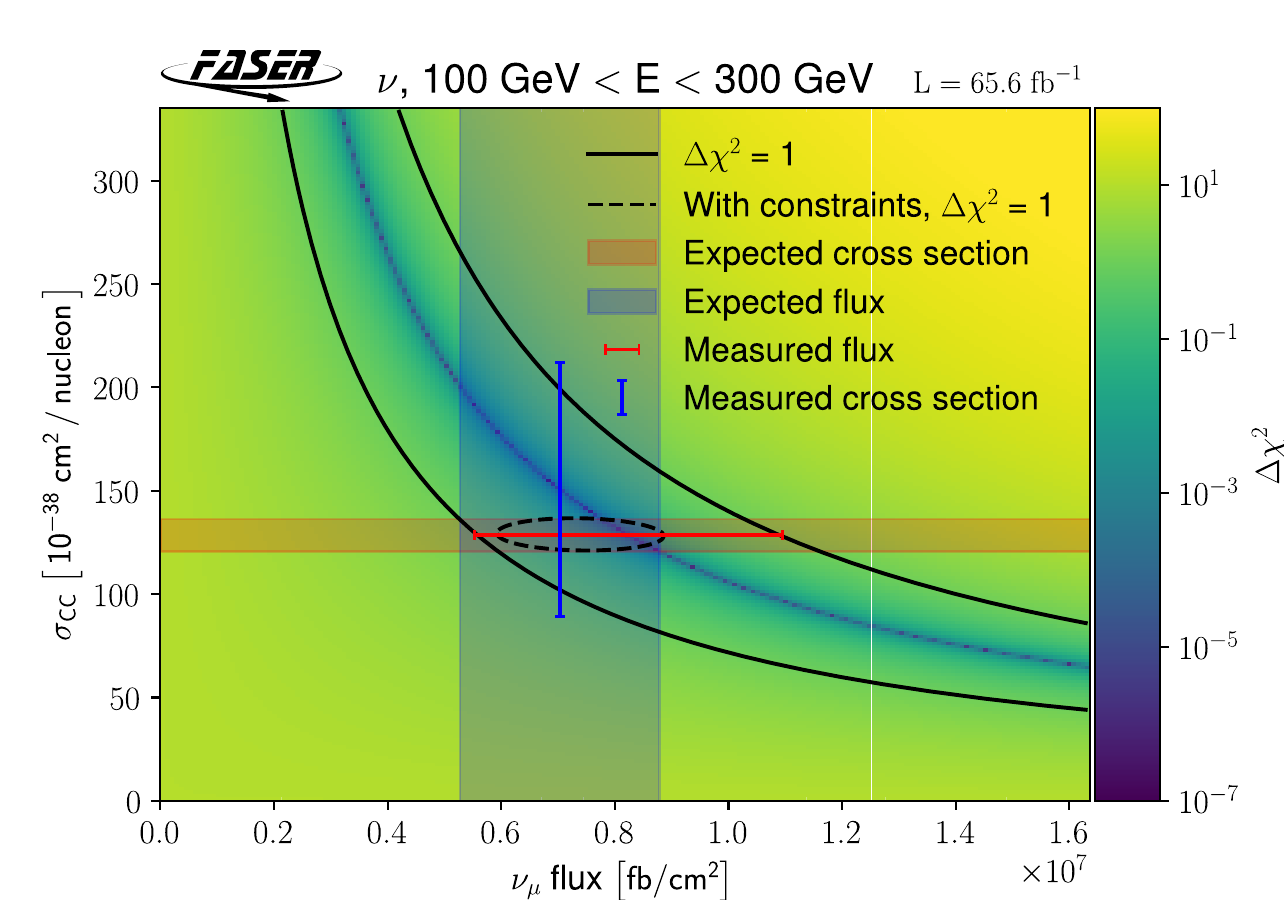}\hspace*{3mm}%
  \includegraphics[width=0.32\textwidth]{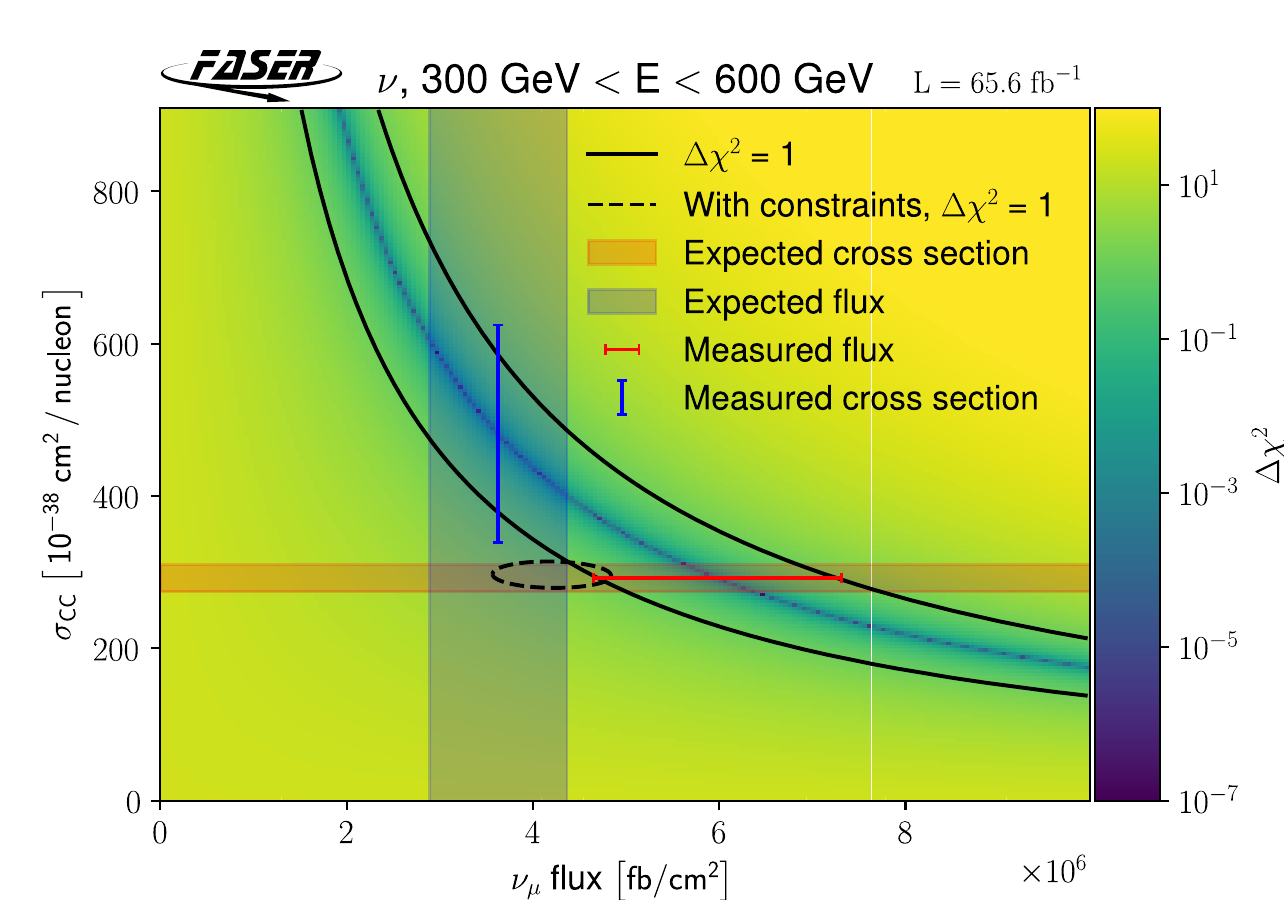}\hspace*{3mm}%
  \includegraphics[width=0.32\textwidth]{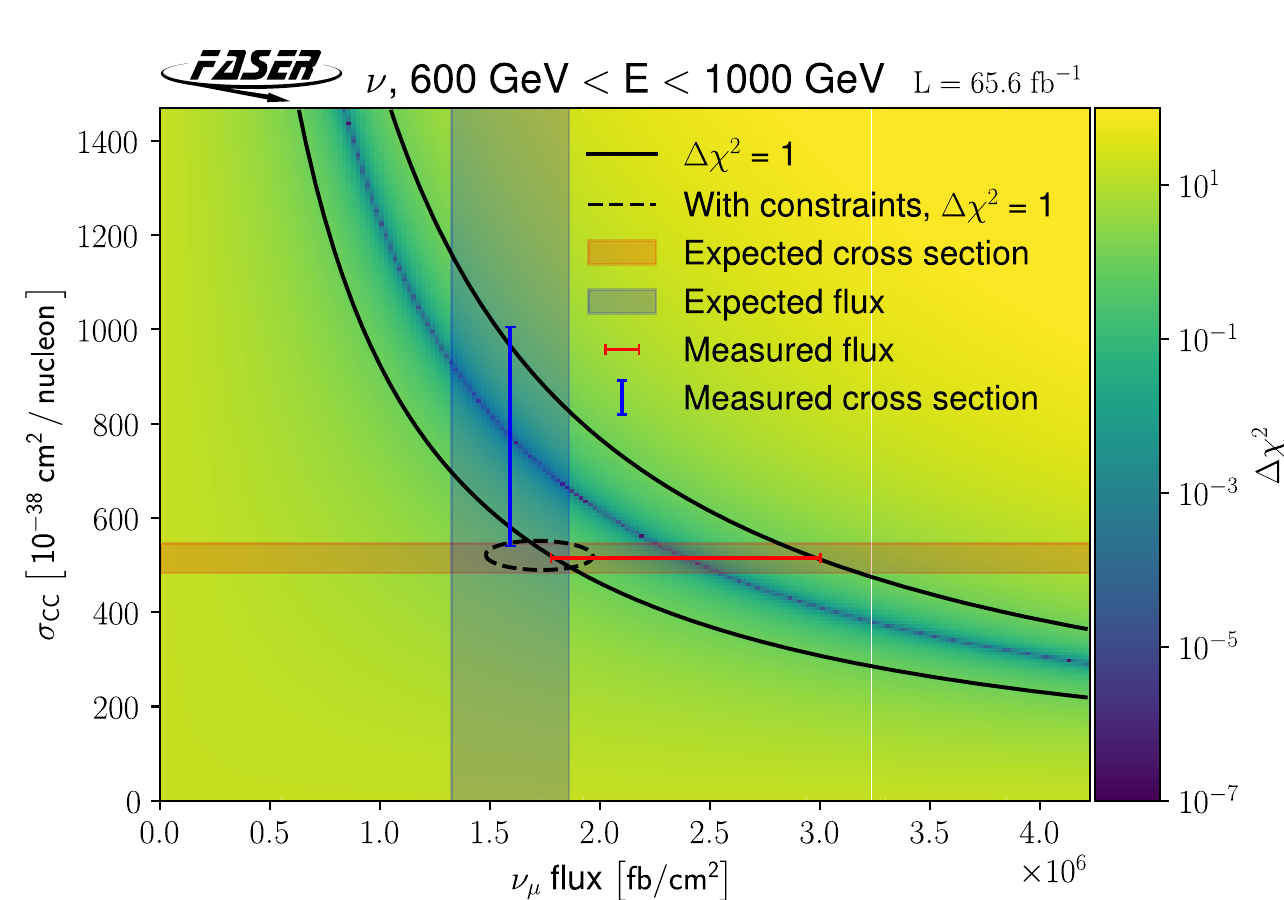}
  \includegraphics[width=0.32\textwidth]{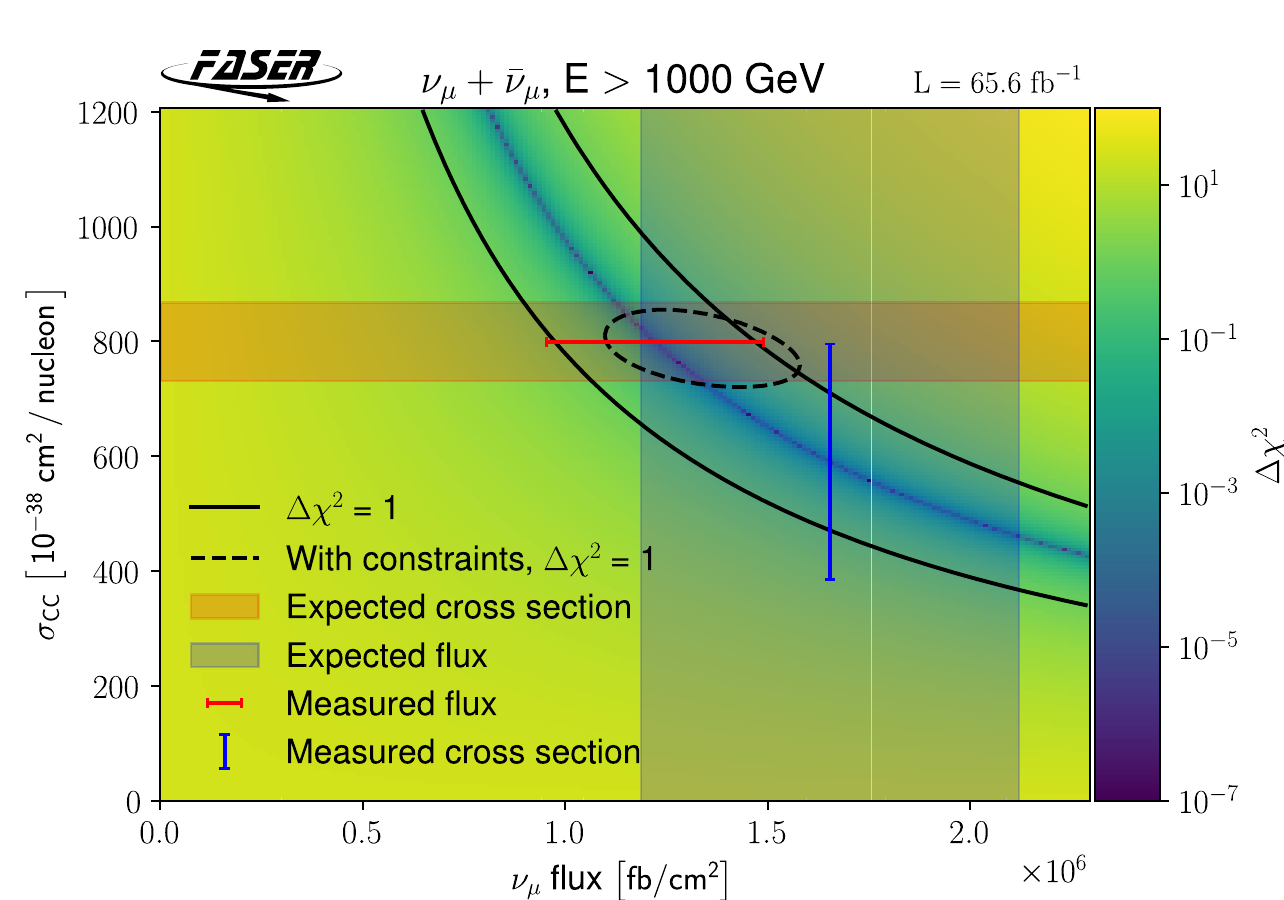}\hspace*{3mm}%
  \includegraphics[width=0.32\textwidth]{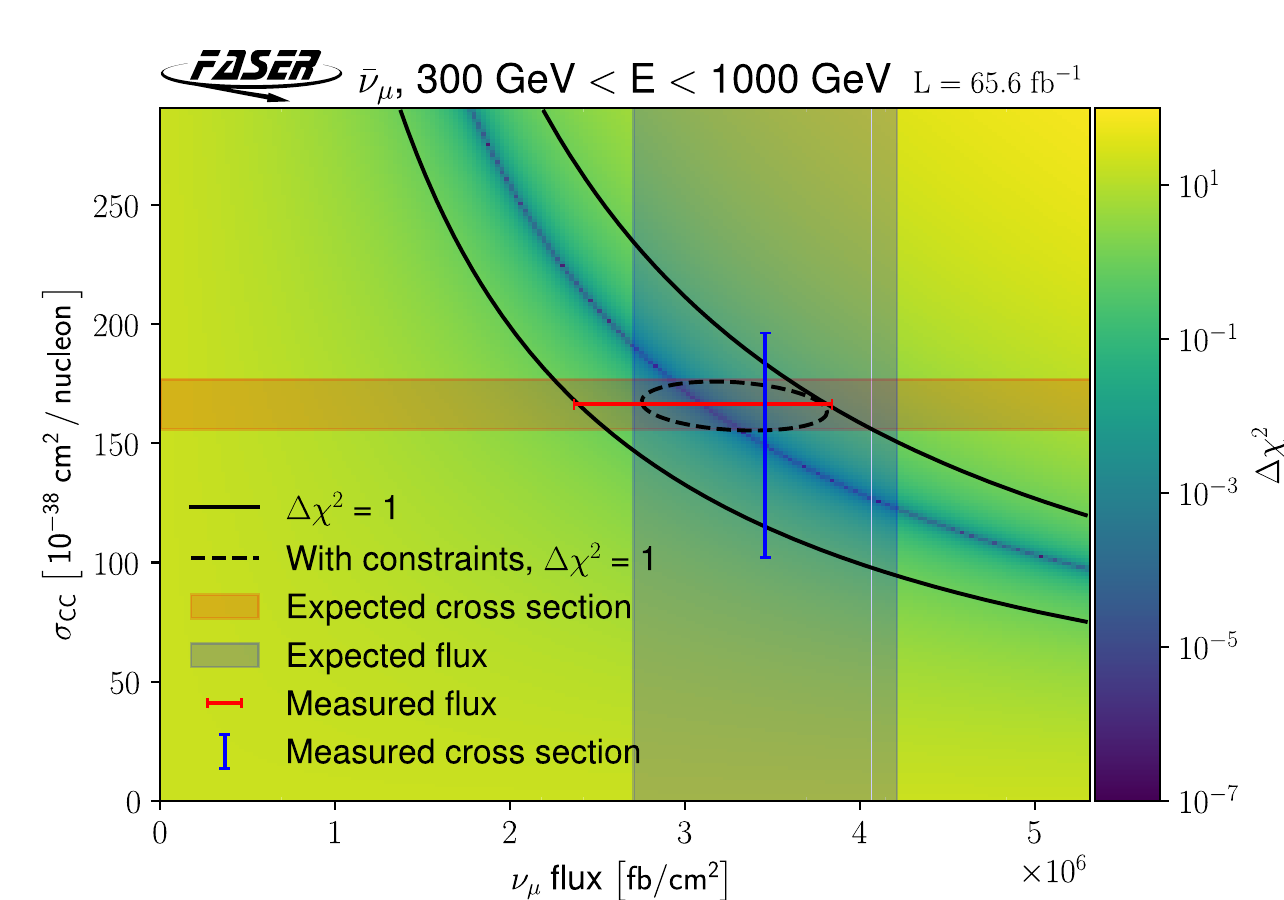}\hspace*{3mm}%
  \includegraphics[width=0.32\textwidth]{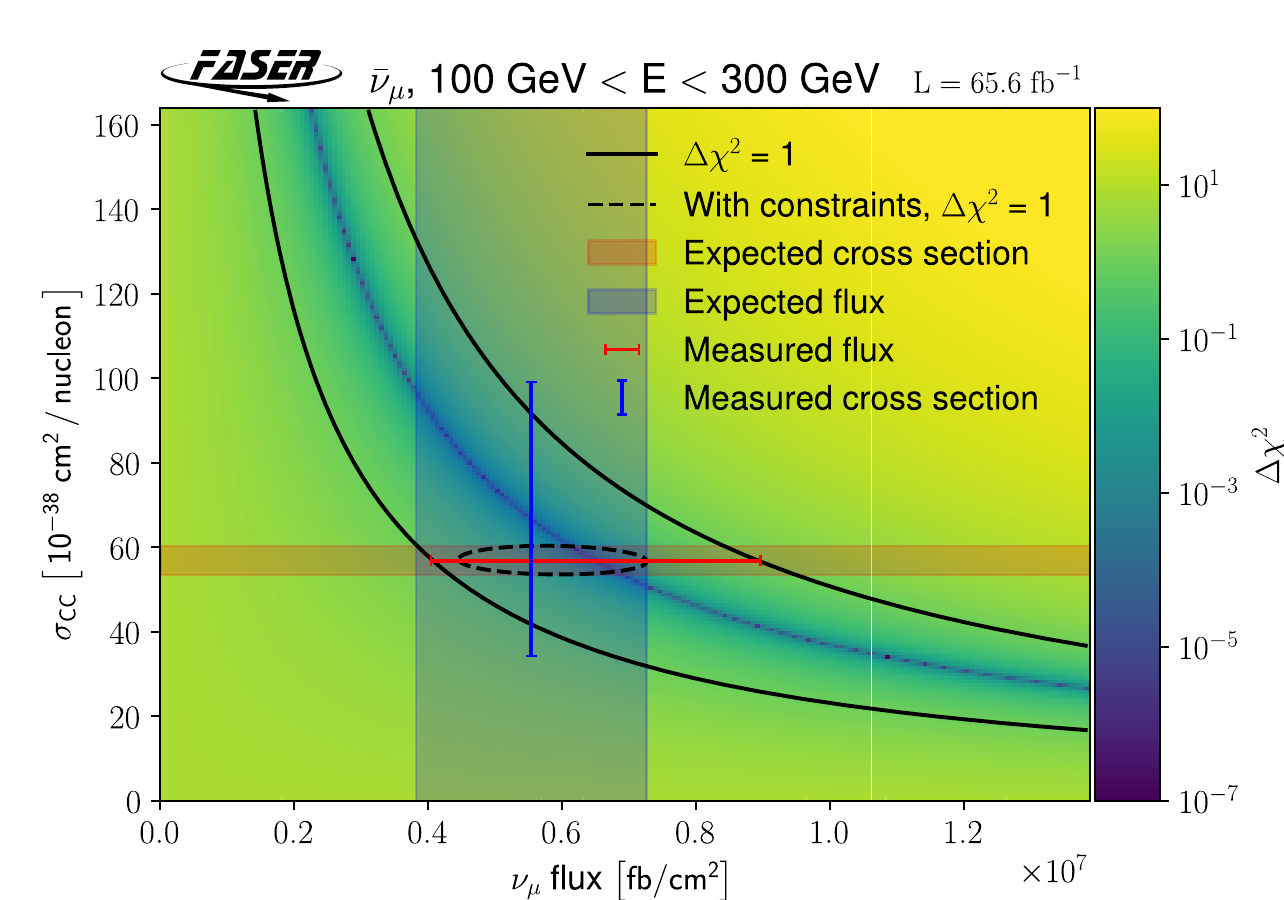}
  \vspace*{-0.15in}
  \caption{
    The contour plots of cross section versus flux for each $-L/E_{\Pnu}$ bin.
    The blue bands and error bars show the expected flux and the corresponding measured cross section.
    And the red bands and error bars show the expected cross section and the corresponding measured flux.
    The black, dashed ellipses show the best fit result when constraining the cross section and flux with the expected values.
  }
  \label{fig:cross-section-flux-contour}
\end{figure*}

To simultaneously measure the flux, $\phi^{k}$, and cross section, $\sigma_{\mathsf{CC}}^k$, in each $-L/E_{\Pnu}$-bin $k$, we minimize the following $\chi^2$ function:
\begin{equation}
  \chi^2_k = \frac{\left(\phi^k \cdot \sigma_{\mathsf{CC}}^k \cdot \rho_T \cdot L - n_{\Pnu}^k \right)^2}{(\delta_n^k)^2},
\end{equation}
with $n_{\Pnu}^k$ and  $\delta_n^k$ denoting the unfolded number and uncertainty of neutrino interactions in bin $k$.
We also perform a constrained fit using the expected flux and cross section  values from simulation
\begin{equation}
  \chi^2_{k, \ \text{const.}} = \frac{\left(\phi^k \cdot \sigma_{\mathsf{CC}}^k \cdot \rho_T \cdot L - n_{\Pnu}^k \right)^2}{(\delta_n^k)^2} + \frac{\left(\phi^{k} - \phi_{\text{sim.}}^k \right)^2}{(\delta_{\phi, \ \text{sim.}}^k)^2} + \frac{\left(\sigma_{\mathsf{CC}}^{k} - \sigma_{\text{sim.}}^k \right)^2}{(\delta_{\sigma, \ \text{sim.}}^k)^2} 
\end{equation}
\Cref{fig:cross-section-flux-contour} compares the measured flux and cross section described above with the results of the simultaneous fits.

\end{document}

%% file: authorlist.tex
\author{Roshan Mammen Abraham\,\orcidlink{0000-0003-4678-3808}}
\affiliation{Department of Physics and Astronomy, University of California, Irvine, CA 92697-4575, USA}

\author{Xiaocong Ai\,\orcidlink{0000-0003-3856-2415}}
\affiliation{School of Physics, Zhengzhou University, Zhengzhou 450001, China}

\author{John Anders\,\orcidlink{0000-0002-1846-0262}}
\affiliation{CERN, CH-1211 Geneva 23, Switzerland}

\author{Claire Antel\,\orcidlink{0000-0001-9683-0890}}
\affiliation{D\'epartement de Physique Nucl\'eaire et Corpusculaire, University of Geneva, CH-1211 Geneva 4, Switzerland}

\author{Akitaka Ariga\,\orcidlink{0000-0002-6832-2466}}
\affiliation{Albert Einstein Center for Fundamental Physics, Laboratory for High Energy Physics, University of Bern, Sidlerstrasse 5, CH-3012 Bern, Switzerland}
\affiliation{Department of Physics, Chiba University, 1-33 Yayoi-cho Inage-ku, 263-8522 Chiba, Japan}

\author{Tomoko Ariga\,\orcidlink{0000-0001-9880-3562}}
\affiliation{Kyushu University, Nishi-ku, 819-0395 Fukuoka, Japan}

\author{Jeremy Atkinson\,\orcidlink{0009-0003-3287-2196}}
\affiliation{Albert Einstein Center for Fundamental Physics, Laboratory for High Energy Physics, University of Bern, Sidlerstrasse 5, CH-3012 Bern, Switzerland}

\author{Florian~U.~Bernlochner\,\orcidlink{0000-0001-8153-2719}}
\affiliation{Universit\"at Bonn, Regina-Pacis-Weg 3, D-53113 Bonn, Germany}

\author{Tobias Boeckh\,\orcidlink{0009-0000-7721-2114}}
\affiliation{Universit\"at Bonn, Regina-Pacis-Weg 3, D-53113 Bonn, Germany}

\author{Jamie Boyd\,\orcidlink{0000-0001-7360-0726}}
\affiliation{CERN, CH-1211 Geneva 23, Switzerland}

% ...
